\documentclass{aa}

\usepackage{graphicx}   
\usepackage{amsmath}    
\usepackage{amssymb}    
\usepackage{multirow}

\usepackage{float}
\usepackage{lscape}

\usepackage{txfonts}
\usepackage{lipsum}
\usepackage{placeins}

\usepackage{txfonts}
\usepackage{hyperref}
\hypersetup{
   colorlinks=true,
   linkcolor=blue,
   citecolor=blue,
   urlcolor=blue
}
\begin{document}

   \title{Detectability of continuous gravitational waves from planetary-mass companions orbiting compact stars}

    \titlerunning{GWs from planetary-mass companions orbiting compact stars}

   \author{Abdusattar Kurban\inst{1,2,3}
    \and
    Xia Zhou\inst{1,2,3}
    \and
    Na Wang\inst{1,2,3}
    \and
    Yong-Feng Huang\inst{4,5}
    \and
    Wenming Yan\inst{1,2,3}
    \and
    Jianping Yuan\inst{1,2,3}
    \and
    Ali Esamdin\inst{3}
    \and
    Yu-Bin Wang\inst{6}
    \and
    Zhigang Wen\inst{1,2,3}
    \and
    Rai Yuen\inst{1,2,3}
}

\institute{State Key Laboratory of Radio Astronomy and Technology, Xinjiang Astronomical Observatory, CAS, 150 Science 1-Street, Urumqi, Xinjiang, 830011, P. R. China\\
    \email{akurban@xao.ac.cn (AK)}
    \and
    Xinjiang Key Laboratory of Radio Astrophysics, Urumqi, Xinjiang, 830011, P. R. China
    \and
    Xinjiang Astronomical Observatory, Chinese Academy of Sciences, Urumqi, Xinjiang, 830011, P. R. China
    \and
    School of Astronomy and Space Science, Nanjing University, Nanjing 210023, P. R. China
    \and
    Key Laboratory of Modern Astronomy and Astrophysics (Nanjing University), Ministry of Education, Nanjing 210023, P. R. China
    \and
    School of Physics and Electronic Engineering, Sichuan University of Science \& Engineering, Zigong 643000, P. R. China
    }

   \date{Received Month xx, yyyy; accepted Month xx, yyyy}

  \abstract
{Binary systems with ultrashort-period planetary-mass companions
are expected to radiate continuous gravitational waves (GWs). However, earlier
studies found that the detectability of such systems by the Laser
Interferometer Space Antenna (LISA) is unlikely. In this study, we investigate the
detectability of GWs from planetary-mass companions
orbiting pulsars (PSRs) or white dwarfs (WDs) whose fundamental
parameters, essential for calculating GW properties, have been
measured. We compare the GW signals from our sample with the
sensitivity curves of space-based GW detectors. We find
that fourteen sources achieve a signal-to-noise ratio (\(\text{S/N}\)) of \(\gtrsim 5\)
within four years of observations. Among these, three sources have
PSR primaries (2S 0918-549 b, 4U 0513-40 b, and 4U 1543-62), and
eleven systems possess WD primaries (BW Scl b, CP Eri b, CR Boo b, EF Eri b,
GP Com b, GW Lib b, SDSS J0926+3624 b, SDSS J1507+5230 b, SMSS J1606-1000 b,
SRGeJ0453 b, and WZ Sge b). We note that their detectability is less probable
with near-term missions such as LISA, TianQin, and Taiji. Nevertheless, they could
be detected by more advanced, future-generation observatories,
such as the Deci-hertz Interferometer Gravitational wave Observatory
(DECIGO) and the Big Bang Observer (BBO). This offers the potential to
investigate the formation and evolution of ultrashort-period
planetary-mass companions around compact stars through joint GW
and electromagnetic surveys.}


   \maketitle

\section{Introduction} \label{sec:intro}

The direct detection of gravitational waves (GWs) from binary black hole mergers \citep{2016PhRvL.116f1102A}
and a binary neutron star inspiral \citep{Abbott2017PhRvL.119p1101A} by ground-based GW detectors operating
in high-frequency bands has opened a new window for observing the Universe.
Space-based GW detectors such as the Laser Interferometer Space Antenna \citep[LISA;][]{Amaro-Seoane2017arXiv170200786A},
TianQin \citep{Luo2016CQGra}, and Taiji \citep{Ruan2020IJMPA} aim to detect GW signals in the milli-hertz frequency
band, which are inaccessible to ground-based detectors. LISA is an approved mission with
well-defined design sensitivity, whereas TianQin and Taiji are in the advanced stages of planning.
The scientific objectives of these instruments encompass the investigation of various compact binary
systems and the stochastic GW background of astrophysical origin \citep{Ruan2020IJMPA,Huang2020PhRvD,Amaro-Seoane2023LRR}.

In addition to the aforementioned instruments, proposed far-future space-based GW
detectors such as the Deci-hertz Interferometer Gravitational wave Observatory \citep[DECIGO;][]{Kawamura2006CQGra}
and the Big Bang Observer \citep[BBO;][]{Harry_2006} are designed to investigate events within the deci-hertz
range of GW frequencies. The projected sensitivities of these two conceptual detectors are several orders of
magnitude superior to those of LISA, TianQin, and Taiji, aiming to explore the early Universe and
compact objects \citep{Cutler2006PhRvD,Yagi2011PhRvD}.
The studies based on these space-based GW instruments are expected to provide a more comprehensive
understanding of the formation and evolution of compact binary systems and to trace the merger
history of compact objects through GW astronomy.

Ultracompact binaries with orbital periods less than 60 minutes have been proposed as potential GW
sources for the LISA mission due to their strong GW signals \citep{Kupfer+2018,Burdge2020ApJ}.
These systems consist of a white dwarf (WD) or neutron star (NS) primary and a compact helium-star, WD,
or NS secondary. \citet{Kupfer+2018} argued that, assuming an optimistic 10-year mission for LISA,
some ultracompact binaries can produce a signal-to-noise ratio (\(\text{S/N}\)) greater than 5; they are referred
to as LISA verification binaries and include AM Canum Venaticorum (AM CVn) systems, double WDs,
subdwarf B-star systems, and ultracompact X-ray binaries (UCXBs). Subsequent studies have
shown that a subclass of UCXBs is detectable by LISA with a considerable \(\text{S/N}\)
\citep{Chen2020ApJ,2021MNRAS.503.2776Y,Suvorov2021MNRAS,Chen2025ApJ}.

It has been proposed that some special planetary-mass companions orbiting pulsars (PSRs)
could serve as potential GW sources for upcoming GW detectors. For instance, small planetary-mass
objects in close orbits around PSRs might belong to strange quark matter planetary systems
\citep{Huang2017ApJ, Kuerban2020ApJ, Zhang2024FrASS}. The strong GW bursts produced during the final
inspiral stage of such systems within our Galaxy could potentially be detected by future ground-based
GW detectors \citep{Geng2015ApJ_a, Kuerban2019AIPC, Zhang2024MNRAS}.

Previous studies have highlighted that continuous GWs from binary systems with ultrashort
period planetary-mass companions, such as GP Com b, could be particularly interesting targets for
the LISA mission \citep{Cunha2018MNRAS}. However, \citet{Wong2019MNRAS} demonstrated that observing
this system for four years would not produce an \(\text{S/N}\) sufficient for detection by LISA,
suggesting that detecting such systems is less likely. In recent years, the number of confirmed 
extrasolar planetary-mass companions has increased significantly. Additionally, except for
LISA, other space-based GW detectors with different sensitivities, such as TianQin, Taiji,
DECIGO, and BBO, are also planned. Based on the sensitivity characteristics of these detectors,
we probe the gravitational-wave signatures of short-period planetary-mass companions
orbiting compact stars using the most up-to-date data.

The structure of this paper is organized as follows. The
theoretical GW signal model is presented in Section
\ref{sec:signal_model}. The parameters of our
planetary-mass companions orbiting compact stars are
given in Section \ref{sec:data}. The characteristics of the GW
signal produced by our selected samples are presented in Section
\ref{sec:results}. Finally, Section \ref{sec:concl_and_discus}
presents our conclusions and some brief discussions.

\section{GW signal model} \label{sec:signal_model}

For a binary system in a circular orbit composed of a primary mass \( m_1 \) and
a companion mass \( m_2 \), the power of emitted GW radiation can be expressed as \citep{Peters+1963}
\begin{equation} \label{eq:edot}
    P = \frac{32}{5} \frac{G^{7 / 3}}{c^{5}}\left(2 \pi f_{\mathrm{orb}} \mathcal{M}_{c}\right)^{10 / 3},
\end{equation}
where $G$ is the gravitational constant, $c$ is the speed of light,
$\mathcal{M}_c = (m_1 m_2)^{3/5}/(m_1 + m_2)^{1/5}$ is the chirp mass, and \(f_{\rm orb}\)
is the orbital frequency related to the orbital period by the equation
\(f_{\rm orb} = 1/P_{\rm orb}\). For a source at a luminosity distance $D_L$, the strain
amplitude of GW is \citep[e.g.,][]{Wagg2022ApJS}
\begin{equation} \label{eq:strain}
    h_{0} = \frac{8}{\sqrt{5}} \frac{(G \mathcal{M}_c)^{5/3}}{c^4 D_L} (2\pi f_{\rm orb})^{2/3}.
\end{equation}
The GW frequency can be estimated by
\begin{equation} \label{eq:f_gw}
    f_{\rm gw} = 2 f_{\rm orb}.
\end{equation}
Its change rate due to the GW radiation is
\begin{equation} \label{eq:fdot}
    \dot{f}_{\rm gw} = \frac{96}{5 \pi} \frac{(G \mathcal{M}_c)^{5/3}}{c^5} (2 \pi f_{\rm orb})^{11/3}.
\end{equation}
The GW signal from a binary system spends approximately $\Delta{T} = f_{\rm gw} / \dot{f}_{\rm gw}$
seconds in the vicinity of a frequency $f_{\rm gw}$, which leads to the signal accumulating at the frequency
$f_{\rm gw}$ \citep{Finn2000}. The accumulated signal represents the characteristic strain amplitude,
$h_{c}$, which can be related to $h_{0}$ as \citep[e.g.,][]{Finn2000, Moore2015}
\begin{equation} \label{eq:char_strain}
    h_{c} = h_{0} \sqrt{N_{\rm cycle}} = h_{0} \sqrt{\frac{f_{\rm gw}^2}{\dot{f}_{\rm gw}}} = h_{0} \sqrt{f_{\rm gw} \Delta{T}}.
\end{equation}
A binary system can be regarded as a stationary source when
$ \dot{f}_{\rm gw} < 1/T_{\rm obs}^2 $, where $T_{\rm obs}$ is the observation duration.
For such systems with $T_{\rm obs} < \Delta{T}$, the characteristic
strain can be estimated as
\begin{equation} \label{eq:char_strain_tobs}
    h_{c} = h_{0} \sqrt{f_{\rm gw} T_{\rm obs}}.
\end{equation}
Let us consider the noise power spectral density for a GW detector, denoted as \( S_{\rm n}(f_{\rm gw}) \).
The noise characteristic amplitude corresponding to \( h_{c} \) can be expressed as
\( S_{h} = \sqrt{f_{\rm gw} S_{\rm n}(f_{\rm gw})} \). In the case of a stationary and circular
binary system, the fully averaged \(\text{S/N}\) (acquired by averaging the \(\text{S/N}\)
across position, inclination, and polarization; \citet{Wagg2022ApJS}) is presented as
\begin{equation} \label{eq:S/N_stat_circ}
    {S/N} = \frac{h_{c}}{S_{h}} = \sqrt{\frac{h_{0}^2 T_{\rm obs}}{S_{\rm n}(f_{\rm gw})}}.
\end{equation}
In subsequent sections, the GW characteristics of our sample are calculated using these equations.

\section{Sample selection} \label{sec:data}

The data of the binary systems with planetary-mass
companions considered in this work were taken from the
Exoplanetary Systems Encyclopaedia catalog
\footnote{\url{https://exoplanet.eu/catalog/}}\citep{Schneider2011AA}.
To date, more than 7000 exoplanets confirmed through
electromagnetic observations have been cataloged in this database.
Computing the GW properties of binary systems hinges on
fundamental parameters such as companion mass, primary mass, orbital
period (or semimajor axis), and our distance to the source.
Therefore, we selected the sources for which these parameters
have been documented in the catalog. We are particularly
interested in the sources whose GW frequency lie within the
frequency range of space-based GW instruments.
Planetary-mass companions with orbital periods $P_{\rm
orb} < 0.07$ days are expected to produce continuous GWs at this
frequency range and can be potential targets for space-based GW
detectors. Interestingly, the systems selected in this
manner are specifically found to consist of planetary-mass objects
orbiting around PSRs and WDs (i.e., PSR systems and WD systems).
The details of these systems are listed in Table \ref{tab:table_data}.

Note that the classification of these companions in our samples as
planets is still a matter of ongoing debate. This is because they exhibit
a mass range extending from approximately 1 to 80 Jupiter mass (\(M_{\rm Jup}\)).
Objects within this mass range could be either brown dwarfs or giant planets. Brown
dwarfs typically possess masses that fall between the deuterium fusion threshold of
13 \(M_{\rm Jup}\) \citep{Spiegel2011ApJ} and the hydrogen fusion threshold of
80 \(M_{\rm Jup}\) \citep{Grieves2021AA}. There has been debate regarding the upper
mass boundary for planets, which could be 25 \(M_{\rm Jup}\) \citep{Schneider2011AA}
or 60 \(M_{\rm Jup}\) \citep{Hatzes2015ApJ}. It has also been posited that
42.5 \(M_{\rm Jup}\) represents the transitional boundary between giant planets
and brown dwarfs \citep{Ma2014MNRAS}.

\section{Results}\label{sec:results}

The GW characteristics of the samples selected in Section
\ref{sec:data} were calculated using the signal model presented in
Section \ref{sec:signal_model}. Our calculations were
performed under the assumption that these systems are stationary
and their orbits are circular. First, we verified the
stationarity by checking the condition $\dot{f}_{\rm gw} <
1/T_{\rm obs}^2$. As shown in Table \ref{tab:table_data}, $\dot{f}_{\rm gw} \leq
3.81\times10^{-18}$ Hz s$^{-1}$ for all samples, which is smaller
than $1/T_{\rm obs}^2 = 1.57\times10^{-17}$ Hz s$^{-1}$ in the
case of the longest observation time considered in this work,
$T_{\rm obs} = 8$ yr. Additionally, for our samples, the numerical
results show that $\Delta T = f_{\rm gw} / \dot{f}_{\rm gw} \geq
1.63\times10^{7}$ yr, which is significantly longer than $T_{\rm
obs}$. Therefore, our stationarity assumption is valid for the
calculations in this work. As for the eccentricity ($e$), this parameter
is unfortunately unavailable for most systems in the catalog or
the relevant literature, except for NGC 6440 X-2 b. For
eccentric binary systems, the dominant GW frequency can be
evaluated by \(f_{\rm gw}=n_{\rm peak}(e)f_{\rm orb}\), where
\(n_{\rm peak}(e) \approx 2\) when \(e \lesssim 0.17\)
\citep{Hamers2021RNAAS}. We notice that NGC 6440 X-2 b has a
relatively small eccentricity of \(e = 0.07\). In fact, for
compact binaries with a small orbital period, the eccentricity is
likely to be generally small because the orbit tends to be
circularized. Hence, we simply took \(n_{\rm peak}(e) = 2\)
and adopted \(f_{\rm gw}=2f_{\rm orb}\) in our calculations. Nevertheless,
it should be noted that for sources with an eccentricity of \(e >
0.17\), the influence of eccentricity on GW properties should
be considered, as in \citet{Wagg2022ApJS}. Below, we present our
numerical results.

\subsection{Characteristic strain}\label{sec:hc}

Gravitational wave frequency was calculated using Equation (\ref{eq:f_gw}),
while the characteristic strain $h_{c}$ was calculated using Equation
(\ref{eq:char_strain_tobs}). The full possible range of \(h_{c}\) 
allowed by the parameter uncertainties was evaluated in a conservative
approach. The upper and lower bounds of \(h_{c}\) characterizing this
potential range were determined by the upper and lower bounds of the strain
amplitude \(h_{0}\), which were calculated by considering the upper and lower
bounds of \(D_L\) and \(\mathcal{M}_c\).

\begin{figure}[tbh!]
    \centering
    \includegraphics[width=0.45\textwidth]{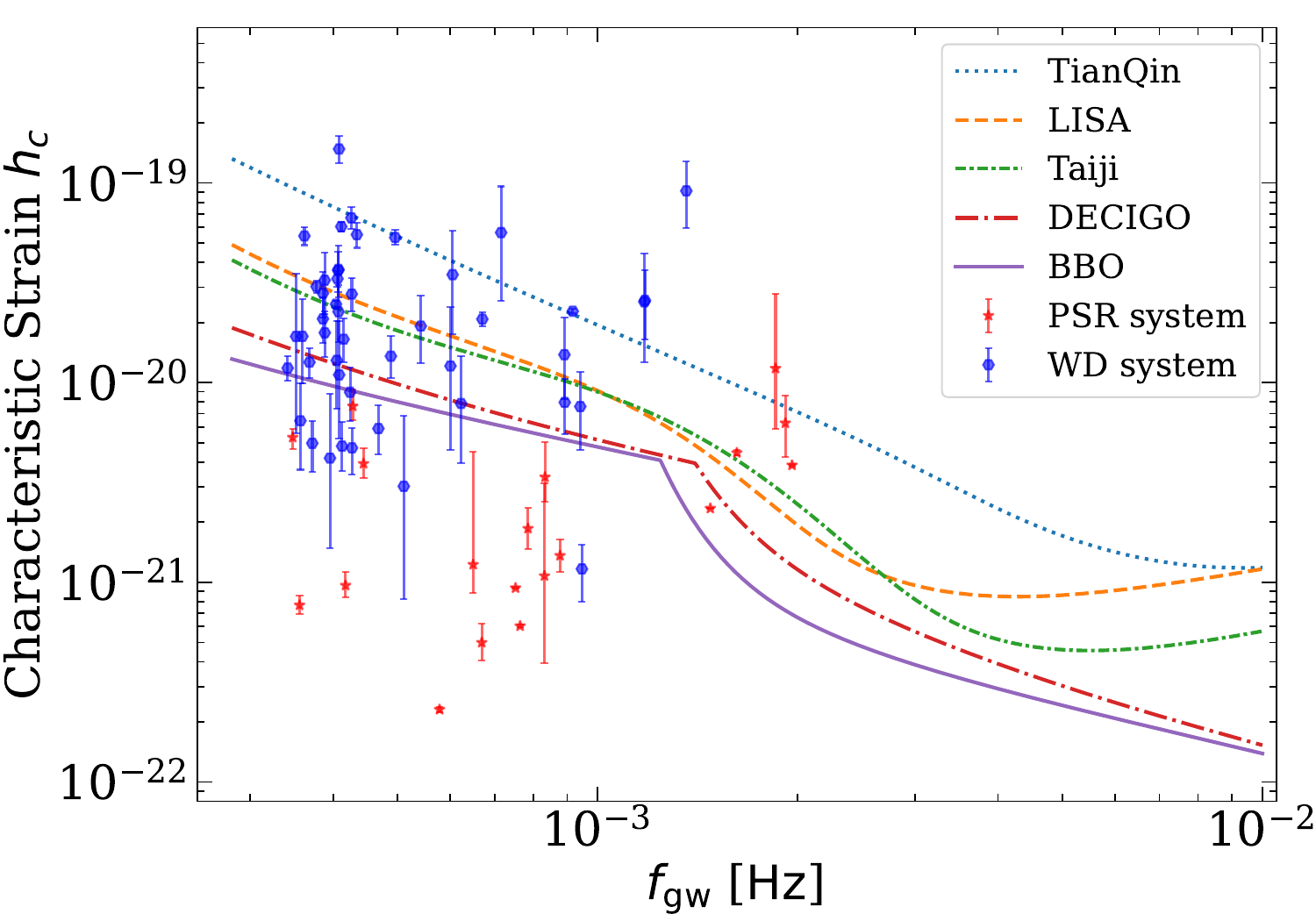}
    \caption{Comparison of the characteristic strain of our sample with \(\sqrt{f_{\rm gw} S_{\rm n}(f_{\rm gw})}\) for \( T_{\rm obs} = 4 \) yr of observations, where \( S_{\rm n}(f_{\rm gw}) \) represents the noise power spectral densities for TianQin \citep[dotted line;][]{Huang2020PhRvD}, LISA \citep[dashed line;][]{Robson2019}, Taiji \citep[densely dashdotted line;][]{Liu2023PhRvD}, DECIGO (dot-dashed line), and BBO (solid line) \citep{Yagi2011PhRvD,Sun2023arXiv231207834S,Sun2024AA}. The stars and hexagons denote the PSR systems and WD systems, respectively.}
    \label{fig:fig1}
\end{figure}

The detectability of the GW signals from sources can be roughly assessed by
comparing the characteristic strain of the source with the sensitivity curve
of a detector, as discussed in \citet{Robson2019}. We obtained the sensitivity
curves of LISA \citep{Robson2019} and TianQin \citep{Huang2020PhRvD} using LEGWORK
\citep{Wagg2022ApJS}. For Taiji, the sensitivity curve was calculated following \citet{Liu2023PhRvD}.
The sensitivity curves for DECIGO and BBO were obtained by incorporating the
confusion noise contributions from both Galactic and extragalactic compact binaries
into the instrumental noise estimates \citep{Yagi2011PhRvD,Yagi2017PhRvD}. This
synthesis was performed following the methodology detailed in
\citet{Sun2023arXiv231207834S} and \citet{Sun2024AA}.

The GW signals from our samples, observed over a period of $ T_{\rm obs} = 4 $ yr,
were compared with the sensitivity curves of these detectors in the frequency versus
characteristic strain plane, as illustrated in Fig. \ref{fig:fig1}. The figure
clearly shows that only a small fraction of our sample exhibits signals that lie
above the sensitivity curves of the TianQin, LISA, and Taiji detectors. In stark
contrast, a substantial population of sources lies above the sensitivity curves
of DECIGO and BBO. However, the actual detectability of these sources, despite
their positions above the noise curves, must be assessed based on their \(\text{S/N}\) values.

\subsection{Signal-to-noise ratio}\label{sec:S/N}

For a source to be considered detectable, its S/N must exceed a certain threshold.
For monochromatic systems, this threshold is typically set at \(\text{S/N} = 5\)
\citep{Kupfer+2018}. Here, we define \(\text{S/N}_{\rm thr} = 5 \) as the
detection threshold. Sources satisfying \(\text{S/N} \geq \text{S/N}_{\rm thr} \)
are deemed detectable.

\begin{figure}
    \centering
    \includegraphics[width=0.49\textwidth]{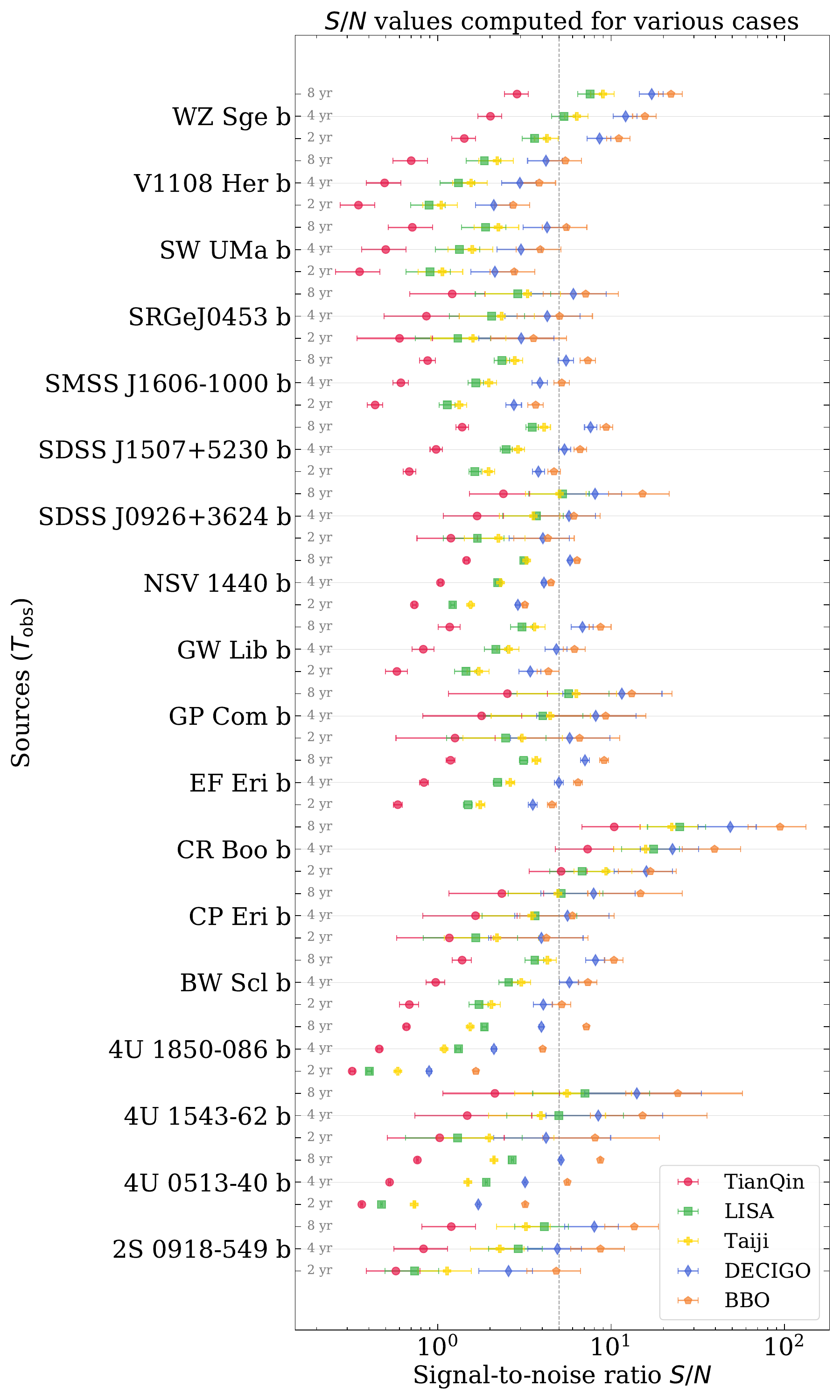}
    \caption{\(\text{S/N}\) values of the promising sources for various detectors and observation durations. The circles, squares, filled pluses, rhombuses, and pentagons represent the TianQin, LISA, Taiji, DECIGO, and BBO results, respectively. The data points situated on the horizontal gray line corresponding to each source signify the \(\text{S/N}\) values calculated for $T_{\rm obs} = 4$ yr. The \(\text{S/N}\) values calculated for $T_{\rm obs} = 2$ yr and $T_{\rm obs} = 8$ yr are shown in the lower and upper sides of the horizontal gray line, respectively. The vertical dashed gray line denotes $\text{S/N}_{\rm thr} = 5$. Sources with $\text{S/N} \gtrsim \text{S/N}_{\rm thr}$ are regarded as detectable. Sources with $\text{S/N} \geq 1$ indicate that the GW signals lie above the sensitivity curves of the corresponding detectors.}
    \label{fig:fig2}
\end{figure}

For monochromatic sources, the \(\text{S/N}\) can be estimated from Equation
(\ref{eq:S/N_stat_circ}), which scales with the square root of the
observation time \( T_{\rm obs} \). We calculated the \(\text{S/N}\) for
our samples with respect to various detectors, each with different
\( T_{\rm obs} \). The results are presented in Table \ref{tab:table_snr}.
From the table, it is evident that the \(\text{S/N}\) depends on the
detectors, observation duration, and source properties. The \(\text{S/N}\)
of each source lies in a range defined by the upper and lower
bounds, computed using the upper and lower bounds of \(h_{0}\).
Here we denote the upper, central, and lower \(\text{S/N}\) values
as \(\text{S/N}_{\rm up} \), \(\text{S/N}_{\rm cen} \),
and \(\text{S/N}_{\rm low} \), respectively, which satisfy
\(\text{S/N}_{\rm up} > \text{S/N}_{\rm cen} > \text{S/N}_{\rm low} \). 
In Tables \ref{tab:table_snr} and \ref{tab:table_3}, the potential range of the S/N for
a mission is represented by the \(\text{S/N}_{\rm cen}\) along with its
superscript (defined as \(\text{S/N}_{\rm up} - \text{S/N}_{\rm cen}\)) and
subscript (defined as \(\text{S/N}_{\rm low} - \text{S/N}_{\rm cen}\)). 
For instance, the observation of 4U 1543-62 b by TianQin for
\(T_{\rm obs} = 4\) yr yields an S/N of \(1.48^{+2.01}_{-0.74}\).
The corresponding \(\text{S/N}_{\rm up}\), \(\text{S/N}_{\rm cen}\),
and \(\text{S/N}_{\rm low}\) are 3.49, 1.48, and 0.74, respectively.
In this work, we focused on sources with \(\text{S/N}_{\rm
cen} \gtrsim \text{S/N}_{\rm thr} \) across all considered
configurations. We refrained from further analyzing sources whose
S/N remained below the detection threshold. As shown in the
table, a subset of sources exhibits \(\text{S/N}_{\rm cen} \gtrsim
\text{S/N}_{\rm thr} \). They are promising candidates for GW
detections. Figure \ref{fig:fig2} plots the \(\text{S/N}\) of these promising
sources for various detectors and observation durations.
Below, we analyze the detectability of these promising sources in detail.

\subsubsection{Sources with a PSR primary}\label{sec:psr-primary}

For the sources with a PSR primary considered here -- as depicted in Fig. \ref{fig:fig2} and Table
\ref{tab:table_snr} -- none yield \(\text{S/N}_{\rm cen}
\gtrsim \text{S/N}_{\rm thr}\) for near-term missions such as
TianQin, LISA, and Taiji within an observation duration of \(T_{\text{obs}} \leq 4 \) yr. 
For far-future conceptual detectors such as DECIGO
and BBO, the number of detectable sources markedly increases.
When \(T_{\text{obs}} = 4\) yr,
2S 0918-549 b, 4U 0513-40 b, and 4U 1543-62 are detectable by BBO with
\(\text{S/N}_{\rm low} > \text{S/N}_{\rm thr} \), while DECIGO
detects 4U 1543-62 b with \(\text{S/N}_{\rm cen} > \text{S/N}_{\rm thr}\). 
Among them, 4U 1543-62 b is detectable by BBO within a comparatively
shorter observation duration of \(T_{\text{obs}} = 2 \) yr, with
\(\text{S/N}_{\rm cen} > \text{S/N}_{\rm thr} \). As shown in
Fig. \ref{fig:fig2} and Table \ref{tab:table_snr}, some sets of
configurations yield \(\text{S/N}_{\rm up} > \text{S/N}_{\rm thr} > \text{S/N}_{\rm cen}\),
which indicates marginal detection with the most optimistic S/N values.

\subsubsection{Sources with a WD primary}\label{sec:wd-primary}

For the sources with a WD primary in our samples, the
detectability also depends on the detectors and
\(T_{\text{obs}}\), as shown in Fig. \ref{fig:fig2} and Table
\ref{tab:table_snr}.
For near-term detectors (TianQin, LISA, and Taiji), the
number of detectable sources is relatively low. When
\(T_{\text{obs}} = 4\) yr, CR Boo b and WZ Sge b have signals
with \(\text{S/N}_{\rm cen} \gtrsim \text{S/N}_{\rm thr}\) for
both LISA and Taiji. But, for TianQin, only CR Boo b has such a
strong signal. Nevertheless, an S/N with a lower bound
\(\text{S/N}_{\rm low} < \text{S/N}_{\rm thr}\) can be generated
by CR Boo b for TianQin and by WZ Sge b for LISA. 
As shown in Fig. \ref{fig:fig2} and Table \ref{tab:table_snr}
CP Eri b, GP Com b, and SDSS J0926+3624 b yield
\(\text{S/N}_{\rm up} > \text{S/N}_{\rm thr} > \text{S/N}_{\rm cen} \)
for LISA and Taiji, indicating potential
detectability when the most optimistic S/N values are taken into account.
When \(T_{\text{obs}} = 2\) yr, only CR
Boo b yields \(\text{S/N}_{\rm cen} > \text{S/N}_{\rm thr}\) for
TianQin, LISA, and Taiji. However, for this source, it is worth
noting that \( \text{S/N}_{\rm low} < \text{S/N}_{\rm thr} \) for TianQin and LISA.

For far-future conceptual detectors (DECIGO and BBO), several sources
yield a signal with \(\text{S/N}_{\rm cen} \gtrsim \text{S/N}_{\rm thr}\)
within \(T_{\text{obs}} = 2\) yr or \(T_{\text{obs}} = 4\) yr of observations.
When \(T_{\text{obs}} = 4\) yr, BW Scl b, CP Eri b, CR Boo b, EF Eri b, GP Com b,
SDSS J0926+3624 b, SDSS J1507+5230 b, and WZ Sge b are detectable to both DECIGO
and BBO, while GW Lib b, SMSS J1606-1000 b, and SRGeJ0453 b are promising targets
for BBO only. Among them, the sources reaching the detection threshold
\( \text{S/N}_{\text{thr}} \) under \(T_{\text{obs}} = 2\) yr are CR Boo b,
GP Com b, WZ Sge b (for both DECIGO and BBO), and BW Scl b (for BBO only).

\subsubsection{Potential sources and detectability enhancement}\label{sec:potential}

\begin{table*}[t!]
	\centering
	\tabcolsep=2.0mm
	\renewcommand{\arraystretch}{1.45}
	\small
	\caption{S/N values of five exemplary sources computed for individual and combined networks using TianQin, LISA, and Taiji, with $ T_{\rm obs} = 4 $ yr.}
	\label{tab:table_3}
	\begin{tabular}{lccccccc}
		\hline\hline
		\noalign{\smallskip}
		Source & TianQin & LISA & Taiji & TianQin+LISA & TianQin+Taiji & LISA+Taiji & TianQin+LISA+Taiji \\
		\hline
		\noalign{\smallskip}
		4U 1543-62 b      & $1.48^{+2.01}_{-0.74}$ & $4.99^{+6.67}_{-2.49}$ & $3.93^{+5.34}_{-1.97}$ & $5.20^{+7.06}_{-2.60}$ & $4.20^{+5.70}_{-2.10}$ & $6.35^{+8.62}_{-3.18}$ & $6.52^{+8.85}_{-3.26}$ \\
		CP Eri b          & $1.65^{+1.22}_{-0.83}$ & $3.64^{+2.69}_{-1.83}$ & $3.49^{+2.58}_{-1.76}$ & $3.99^{+2.95}_{-2.01}$ & $3.86^{+2.85}_{-1.94}$ & $5.04^{+3.73}_{-2.54}$ & $5.30^{+3.92}_{-2.67}$ \\
		GP Com b          & $1.79^{+1.26}_{-0.97}$ & $4.03^{+2.83}_{-2.19}$ & $4.45^{+3.13}_{-2.42}$ & $4.41^{+3.10}_{-2.39}$ & $4.79^{+3.37}_{-2.61}$ & $6.00^{+4.22}_{-3.26}$ & $6.26^{+4.41}_{-3.40}$ \\
		SDSS J0926+3624 b & $1.68^{+0.71}_{-0.61}$ & $3.72^{+1.56}_{-1.35}$ & $3.56^{+1.50}_{-1.29}$ & $4.08^{+1.72}_{-1.48}$ & $3.94^{+1.66}_{-1.43}$ & $5.15^{+2.16}_{-1.87}$ & $5.42^{+2.28}_{-1.97}$ \\
		WZ Sge b          & $2.01^{+0.32}_{-0.30}$ & $5.35^{+0.86}_{-0.81}$ & $6.34^{+1.02}_{-0.96}$ & $5.72^{+0.92}_{-0.87}$ & $6.65^{+1.07}_{-1.01}$ & $8.30^{+1.34}_{-1.26}$ & $8.54^{+1.38}_{-1.29}$ \\
		\noalign{\smallskip}
		\hline
	\end{tabular}
\end{table*}

In addition to the systems with \(\text{S/N}_{\rm cen} \gtrsim 5\)
discussed above, sources with \(5 > \text{S/N}_{\rm cen} > 3\) and
those marginally detected with the most optimistic S/N values
(i.e., \(\text{S/N}_{\rm up} \gtrsim \text{S/N}_{\rm thr} > \text{S/N}_{\rm cen}\))
also deserve our attention as potential targets for GW detectors. The detectability
of these sources can be enhanced through two main methods.

First, extending the mission duration can significantly improve the \(\text{S/N}\),
thereby increasing the likelihood of detection. A longer observation period
allows more signal to accumulate, amplifying the \(\text{S/N}\) and bringing these
sources closer to or above the detection threshold.
In our samples, some systems that fail to reach the detection
threshold \(\text{S/N}_{\rm thr}\) for \(T_{\text{obs}} = 4\) yr can
yield an \(\text{S/N}_{\rm cen}\) value higher than 5 after being
observed for \(T_{\text{obs}} = 8\) yr. For systems with a PSR primary,
these include 4U 1543-62 b for LISA and Taiji, 2S 0918-549 b and 4U 0513-40 b
for DECIGO, and 4U 1850-086 b for BBO. Similarly, for systems with a WD
primary, nine sources yield a signal with
\(\text{S/N}_{\rm cen} \gtrsim \text{S/N}_{\rm thr}\) after being observed
for \(T_{\text{obs}} = 8\) yr. They include three systems (CP Eri b, GP Com b,
and SDSS J0926+3624 b) for LISA and Taiji, one system (NSV 1440 b) for both
DECIGO and BBO, three systems (GW Lib, SMSS J1606-1000 b, and SRGeJ0453 b)
for DECIGO, and two systems (SW UMa b and V1108 Her b) for BBO.

Second, the \(\text{S/N}\) values of such potential sources can be substantially
enhanced by network observations. Specifically, the \(\text{S/N}\) of a source
in a joint observation campaign is calculated as the square root of the
sum of the squares of the individual \(\text{S/N}\) values from each detector,
i.e., \(\text{S/N} = \sqrt{\sum \text{S/N}_{i}^{2}}\) \citep{Wu2023PhRvD},
where \(i\) represents the $i$th detector of network observations.
This enhancement is particularly effective when the individual \(\text{S/N}\) values are comparable.
Table \ref{tab:table_3} lists the \(\text{S/N}\) values yielded by individual and combined
detector networks. We see that the combined network of detectors can considerably
enhance the \(\text{S/N}\). For example, independent observations of 4U 1543-62 b by a single
detector of TianQin, LISA, and Taiji over an observation period
of \( T_{\rm obs} = 4 \) yr yield an \(\text{S/N}\) of \(1.48^{+2.01}_{-0.74}\),
\(4.99^{+6.67}_{-2.49}\), and \(3.93^{+5.34}_{-1.97}\), respectively.
In comparison, the network observation attains a \(\text{S/N}\)
of \(5.20^{+7.06}_{-2.60}\), \(4.20^{+5.70}_{-2.10}\), \(6.35^{+8.62}_{-3.18}\),
and \(6.52^{+8.85}_{-3.26}\) for the combined networks TianQin+LISA,
TianQin+Taiji, LISA+Taiji, and TianQin+LISA+Taiji, respectively.
The table likewise indicates that the \( \text{S/N}_{\rm cen} \)
yielded by most network configurations (TianQin+LISA, LISA+Taiji, and TianQin+LISA+Taiji)
exceeds the detection threshold \(\text{S/N}_{\rm thr} \).
Similarly, for systems with a WD primary, the sources CP Eri b, GP Com b, SDSS J0926+3624 b,
and WZ Sge b could serve as potential targets for the combined networks
TianQin+LISA, TianQin+Taiji, LISA+Taiji, and/or TianQin+LISA+Taiji. We also
notice that \( \text{S/N}_{\rm thr} > \text{S/N}_{\rm low} \) holds for the
results obtained from these network observations, suggesting that the parameter
uncertainties still considerably influence detectability. We posit
that the future DECIGO+BBO network has the potential to further enhance
detectability significantly.

Moreover, the detectability of these sources is also influenced by their geometric
properties. The sky position, orbital inclination, and polarization angle of a system
can affect the \(\text{S/N}\) by a factor of a few due to the amplitude modulation caused by the
detector's orbital motion \citep{Huang2020PhRvD,Wagg2022ApJS}. Favorable orientations
(inclination and polarization angle) and/or advantageous sky locations can enhance the
\(\text{S/N}\) sufficiently to make these sources detectable.

\section{Conclusions and discussions} \label{sec:concl_and_discus}

In this study, we investigated the detectability of GWs from planetary-mass objects around compact stars
by various space-based GW detectors. We focused on ultrashort-period systems (\(P_{\rm orb} < 0.07\) days)
with measured fundamental parameters essential for calculating their GW emission properties. Our samples
consisted of systems featuring a PSR primary or a WD primary. We compared the GW signals from our samples to the
sensitivity curves of near-term detectors (TianQin, LISA, and Taiji), and next-generation conceptual
instruments (DECIGO and BBO) in a characteristic strain versus GW frequency diagram and calculated their
\(\text{S/N}\) values. We find that a subset of these systems is detectable by these GW observatories and summarize our main results below.

For systems with a PSR primary, the sources 2S 0918-549 b, 4U 0513-40 b, and 4U 1543-62 yield
a signal with \( \text{S/N}_{\rm cen} \gtrsim 5\) in an observational duration of \(T_{\rm obs} = 4\) yr in
view of next-generation conceptual instruments. In particular, 4U 1543-62 b is a good target for DECIGO, while
2S 0918-549 b, 4U 0513-40 b, and 4U 1543-62 b are promising candidates for BBO.

For systems with a WD primary, eleven sources are identified as potential candidates,
i.e., BW Scl b, CP Eri b, CR Boo b, EF Eri b, GP Com b, GW Lib b, SDSS J0926+3624 b, SDSS J1507+5230 b,
SMSS J1606-1000 b, SRGeJ0453 b, and WZ Sge b. These sources yield a signal with
\(\text{S/N}_{\rm cen} \gtrsim 5\) in \(T_{\rm obs} = 4 \) yr in view of such next-generation
conceptual detectors as DECIGO and BBO. Among these sources, CR Boo b and WZ Sge b can even be
detectable to some near-term GW observatories such as TianQin, LISA, and Taiji.

Planetary-mass objects around compact stars represent a promising class of targets for space-based
GW instruments. However, they are unlikely to be detected by near-term detectors unless the systems have
relatively large chirp masses and are close to us. Nevertheless, they are good targets for next-generation
conceptual instruments. Especially, increasing the observation time and performing network observation by
using more than one mission could significantly increase detectability. Gravitational wave observations of these systems
can augment our comprehension of substellar mass objects near compact stars, which may include
planets, brown dwarfs, or other objects with masses less than \( 80 M_{\rm Jup}\).

On the one hand, low-mass companions of compact stars could evolve into objects whose masses lie below
60 \(M_{\rm Jup}\) \citep{Sengar2017MNRAS}. Such binaries with a WD primary are thought to evolve into
AM CVn systems through a series of evolutionary stages, yet their precise formation pathways remain obscure
\citep{2018MNRAS.476.1663G,2022ApJ...935....9C}. Systems with an NS primary can evolve into UCXBs,
subsequently widening their orbits \citep{Heinke2013ApJ,Chen2020ApJ}. In such systems, NSs may eventually
become millisecond radio PSRs, accompanied by planetary-mass objects \citep{Wang2021MNRAS}.

On the other hand, according to the strange quark matter hypothesis \citep{Witten1984PhRvD},
strange quark matter objects with masses ranging from planet to pulsar may exist, either in a bare
state or covered by a crust of normal matter \citep{Alcock1986ApJ,Kurban2022PhLB}. The orbits of
such planets can be remarkably tight, allowing them to venture extremely close to their host stars
without being torn apart \citep{Huang2017ApJ,Kuerban2020ApJ}.

The behavior and evolution of objects orbiting compact stars vary significantly, depending on their
interior compositions. In strange-quark-matter planetary systems, tidal effects remain negligible until
the final few inspiral orbits, provided the planet is fully bare \citep{Wang2021PhRvD}. Conversely, if
the strange-matter core is covered by a normal-matter crust, the system may instead power electromagnetic
transient phenomena \citep{Geng2021Innov,Zhou2025ApJ}. In contrast, in binaries containing
nondegenerate stars or tight systems with WDs, tidal interactions and mass loss can play crucial roles,
resulting in discernible differences in the GW signal \citep{Amaro-Seoane2023LRR}. The distinct signatures
left in the GWs emitted by binary systems with planetary-mass companions around PSRs can provide
valuable information on the composition of the secondary objects.

Investigating close-in planets around compact stars through GW astronomy offers a complementary avenue for
discovering new ultrashort period planetary-mass companions. A multi-messenger approach, combining GW and
electromagnetic observations, can provide deeper insights into the nature of these systems. Such joint analyses
are expected to yield crucial information about both the formation and the long-term evolution of
ultrashort-period planetary-mass companions orbiting compact stars, with valuable implications for
physics and astrophysics.

\begin{acknowledgements}
We would like to thank the anonymous referee for helpful suggestions that led to a significant improvement of our work.
This study is supported by the National Key R\&D Program of China (Nos. 2022YFC2205202, 2021YFA0718500, 2022YFC2205203),
the National Natural Science Foundation of China (Grant Nos. 12573051, 12288102, 12273028, 12233002, 12273100, 12303053),
the Natural Science Foundation of Xinjiang Uygur Autonomous Region (Nos. 2022D01A363, 2023D01E20),
the Major Science and Technology Program of Xinjiang Uygur Autonomous Region (No. 2022A03013-1),
the Tianshan Talent Training Program of Xinjiang Uygur Autonomous Region, China (Grant Nos. 2023TSYCLJ0053, 2023TSYCTD0013),
the Sichuan Provincial Natural Science Foundation Project (No. 2025ZNSFSC0878),
the talent introduction program of Sichuan University of Science \& Engineering (No. 2024RC15).
Y.F.H. acknowledges the support from the Xinjiang Tianchi Program.
A.K. acknowledges the support from the Tianchi Talents Project of Xinjiang Uygur Autonomous Region.
W.M.Y. is supported by the West Light Foundation of the Chinese Academy of Sciences (No. WLFC 2021-XBQNXZ027).
Z.G.W. is supported by the 2021 project Xinjiang Uygur Autonomous Region of China for Tian Shan elites,
the Youth Innovation Promotion Association of CAS under No. 2023069,
and the Tianshan Talent Training Program (No. 2023TSYCCX0100).
\end{acknowledgements}

\bibliographystyle{aa} 
\bibliography{reference} 

\begin{appendix}

\onecolumn

\section{Observed and derived properties of the investigated sources}

For the investigated sources, the parameters observed electromagnetically and the derivative of the GW frequency ($\dot{f}_{\rm gw}$) are presented in Table \ref{tab:table_data}, and the S/N values of the GW calculated for various detectors and observation durations are listed in Table \ref{tab:table_snr}.

{\small
	\tabcolsep=3.6mm
	\centering
	\renewcommand{\arraystretch}{1.45}
	\begin{longtable}{lccccccc}
		\caption{Parameters of the sources considered in this study.} \label{tab:table_data}\\
		\hline\hline
		\noalign{\smallskip}
		Source name & $m_2$ & $P_{\rm orb}$ & $e$ & $D_{L}$ & $m_1$ & $\dot{f}_{\rm gw}$ & References \\
		& $(M_{\rm jup})$ & (day) &  & (pc) & $(M_{\odot})$ & ($\rm {Hz\,s^{-1}}$) &  \\
		\hline
		\endfirsthead
		\caption{Continued.}\\
		\hline\hline
		\noalign{\smallskip}
		Source name & $m_2$ & $P_{\rm orb}$ & $e$ & $D_{L}$ & $m_1$ & $\dot{f}_{\rm gw}$ & References \\
		& $(M_{\rm jup})$ & (day) &  & (pc) & $(M_{\odot})$ & ($\rm {Hz\,s^{-1}}$) &  \\
		\hline
		\endhead
		\hline
		\endfoot
		\multicolumn{8}{c}{Source with a PSR primary}\\
		\cline{1-8}
		2S 0918-549 b&27.8$\pm$2.6&0.012083&$\dots$&$4243^{+1468}_{-868}$&1.4& 2.08e-18&[1,2]\\
		4U 0513-40 b&47.1&0.011806&$\dots$&$11951^{+134}_{-133}$&1.4& 3.81e-18&[3,4,5]\\
		4U 0614+09 b&14.9&0.035625&$\dots$&$3300^{+1300}_{-2400}$&1.4& 2.12e-20&[4,6,7]\\
		4U 1543-62 b&42.4&0.012500&$\dots$&$3300^{+3300}_{-1900}$&1.4& 2.79e-18&[4,8,9]\\
		4U 1626-67 b&41.9&0.027777&$\dots$&$4500\pm1500$&1.4& 1.47e-19&[4,10,11]\\
		4U 1850-086 b&41.9&0.014292&$\dots$&$7382^{+240}_{-233}$&1.4& 1.69e-18&[4,5,12]\\
		4U 1905+000 b&52.4&0.055417&$\dots$&$8750\pm1250$&1.4& 1.46e-20&[4,13,14]\\
		IGR J17062-6143 b&22.6&0.026368&$\dots$&$7300\pm500$&1.7$\pm$0.3& 1.10e-19&[4,15,16]\\
		M15 X-2 b&35.6&0.015667&$\dots$&$10709^{+96}_{-95}$&1.4& 1.02e-18&[4,17]\\
		MAXI J0911-655 b&29.3&0.030764&$\dots$&$10060^{+112}_{-111}$&1.4& 7.11e-20&[5,18]\\
		NGC 6440 X-2 b&8.0&0.040022&0.07&$8248^{+248}_{-241}$&1.4& 7.67e-21&[5,19]\\
		NGC 6652B b&17.4&0.030278&$\dots$&$9464^{+139}_{-137}$&1.4& 4.49e-20&[4,5,20]\\
		PSR J0636+5129 b&8.5&0.066551&$\dots$&$210^{+30}_{-20}$&1.4& 1.22e-21&[21,22]\\
		PSR J1311-3430 b&12.6$\pm$0.6&0.065000&$\dots$&$3010\pm150$&2.22$\pm$0.1& 2.69e-21&[23]\\
		PSR J1653-0158 b&14.7$\pm$1.0&0.052083&$\dots$&$870^{+70}_{-80}$&2.15$\pm$0.16& 6.92e-21&[23]\\
		XB 1916-053 b&15.6$\pm$1.0&0.034583&$\dots$&$8800\pm1300$&1.4& 2.47e-20&[4,14,24]\\
		XTE J1751-305 b&36.7&0.029460&$\dots$&$7600\pm900$&1.7$\pm$0.3& 1.19e-19&[4,25,26]\\
		XTE J1807-294 b&15.6$\pm$7.5&0.027829&$\dots$&$5500\pm2500$&1.5$\pm$0.5& 5.74e-20&[4,27]\\
		ZTF J1406+1222 Ab&52.4&0.054057&$\dots$&$1140\pm200$&1.4& 1.60e-20&[28]\\
		\cline{1-8}
		\multicolumn{8}{c}{Source with a WD primary}\\
		\cline{1-8}
		AL Com b&49.2$\pm$10.5&0.056668&$\dots$&$523^{+252}_{-149}$&0.9$\pm$0.3& 9.37e-21&[29,30]\\
		ASASSN-16kr b&44.0$\pm$1.0&0.061286&$\dots$&$160.50^{+7.16}_{-6.57}$&0.952$\pm$0.018& 6.54e-21&[31,32]\\
		ASASSN-17jf b&62.9$\pm$8.4&0.056790&$\dots$&$260.70^{+62.55}_{-42.27}$&0.669$\pm$0.031& 9.63e-21&[2,31]\\
		BW Scl b&53.4$\pm$6.3&0.054324&$\dots$&$93.65^{+0.46}_{-0.46}$&0.85$\pm$0.04& 1.14e-20&[33]\\
		CP Eri b&51.3$\pm$21.0&0.019690&$\dots$&$743.44^{+191.84}_{-126.54}$&0.8$\pm$0.1& 4.34e-19&[2,34,35]\\
		CR Boo b&69.1$\pm$23.0&0.017025&$\dots$&$351.64^{+4.60}_{-4.48}$&0.885$\pm$0.215& 1.06e-18&[2,34,36]\\
		CRTS J012059.6+325545 b&49.8$\pm$5.3&0.057145&$\dots$&$330.04^{+54.10}_{-40.74}$&0.586$\pm$0.29& 6.84e-21&[2,37,38]\\
		CRTS J1122-1110 b&14.7$\pm$5.2&0.045300&$\dots$&$701.16^{+930.69}_{-254.66}$&0.83$\pm$0.23& 6.08e-21&[2,39]\\
		DI UMa b&53.4$\pm$11.5&0.054558&$\dots$&$685^{+43}_{-31}$&0.83$\pm$0.17& 1.10e-20&[30,40]\\
		EF Eri b&83.7$\pm$3.0&0.056260&$\dots$&$160.44^{+3.73}_{-3.57}$&0.9& 1.62e-20&[2,38]\\
		EG Cnc b&32.5$\pm$7.3&0.059940&$\dots$&$186.55^{+7.21}_{-6.69}$&1.03$\pm$0.05& 5.55e-21&[2,29,30]\\
		EPIC 212235321 b&57.6$\pm$10.5&0.047370&$\dots$&$386.83^{+26.99}_{-23.68}$&0.47$\pm$0.01& 1.35e-20&[32,41,42]\\
		EZ Lyn b&44.0$\pm$14.7&0.059580&$\dots$&$142.78^{+2.29}_{-2.22}$&0.85$\pm$0.1& 6.72e-21&[43]\\
		GP Com b&26.2$\pm$16.6&0.032338&$\dots$&$72.83\pm0.24$&0.528$\pm$0.027& 2.74e-20&[2,44,45]\\
		GW Lib b&52.4$\pm$7.3&0.053300&$\dots$&$113.05^{+0.79}_{-0.78}$&0.84$\pm$0.02& 1.19e-20&[2,46]\\
		IX Draconis b&58.1$\pm$26.6&0.064800&$\dots$&$752.11^{+23.27}_{-21.91}$&0.7$\pm$0.1& 5.67e-21&[2,38,47]\\
		NSV 1440 b&30.6&0.025233&$\dots$&$324.23^{+8.80}_{-8.35}$&0.65& 9.14e-20&[2,48]\\
		OGLE BLG-DN-7 b&51.3$\pm$12.4&0.059558&$\dots$&$176.51^{+3.11}_{-3.00}$&0.388$\pm$0.045& 4.54e-21&[2,37,38]\\
		OT J1112-3538 b&36.4$\pm$1.2&0.058500&$\dots$&$499.25^{+180.19}_{-104.65}$&0.338$\pm$0.286& 3.16e-21&[2,37,38]\\
		PQ And b&58.7&0.055800&$\dots$&$267.47^{+16.29}_{-14.53}$&0.47$\pm$0.13& 7.51e-21&[2,38,49]\\
		PR Her b&43.0&0.054200&$\dots$&$571.46^{+204.27}_{-119.12}$&0.34& 4.93e-21&[2,37,38]\\
		QZ Lib b&33.5$\pm$12.6&0.064360&$\dots$&$187.36^{+12.43}_{-10.97}$&0.83$\pm$0.23& 3.81e-21&[32,50]\\
		SDSS J0926+3624 b&36.7$\pm$3.1&0.019650&$\dots$&$548.31^{+235.95}_{-126.81}$&0.85$\pm$0.04& 3.28e-19&[32,51]\\
		SDSS J1035+0551 b&54.5$\pm$2.1&0.057000&$\dots$&$195.24^{+11.57}_{-10.34}$&0.94$\pm$0.01& 1.04e-20&[2,52,53]\\
		SDSS J1057+2759 b&45.7$\pm$2.1&0.062800&$\dots$&$343.76^{+49.69}_{-38.55}$&0.8$\pm$0.015& 5.51e-21&[2,54,55]\\
		SDSS J1240-0159 b&32.5&0.025940&$\dots$&$658.02^{+290.47}_{-154.27}$&0.38& 6.05e-20&[2,56]\\
		SDSS J1339+4847 b&55.3&0.057290&$\dots$&$151.08^{+1.48}_{-1.45}$&0.42$\pm$0.02& 5.95e-21&[2,37,38]\\
		SDSS J1507+5230 b&58.7$\pm$1.1&0.046667&$\dots$&$160\pm10$&0.9$\pm$0.01& 2.27e-20&[57]\\
		SDSS J1433+1011 b&57.6$\pm$8.4&0.054241&$\dots$&$233.29^{+8.85}_{-8.22}$&0.8$\pm$0.07& 1.18e-20&[2,58]\\
		SDSS J1730+5545 b&6.0&0.024444&$\dots$&$1232.44^{+573.59}_{-297.07}$&0.6& 1.93e-20&[2,59]\\
		SMSS J1606-1000 b&68.1$\pm$5.2&0.063915&$\dots$&$107.90^{+1.56}_{-1.52}$&0.72$\pm$0.05& 7.10e-21&[2,38,60]\\
		SRGeJ0411 b&41.9&0.067728&$\dots$&$320.10^{+30.01}_{-25.27}$&0.84$\pm$0.07& 3.97e-21&[61,62]\\
		SRGeJ0453 b&46.1$\pm$21.0&0.038250&$\dots$&$234.80^{+9.80}_{-9.05}$&0.85$\pm$0.04& 3.57e-20&[32,62]\\
		SSS J0522-3505 b&44.0$\pm$0.4&0.062193&$\dots$&$823.72^{+298.61}_{-173.11}$&0.76$\pm$0.023& 5.32e-21&[31,62]\\
		SW UMa b&60.8$\pm$11.5&0.056810&$\dots$&$161.29^{+1.58}_{-1.55}$&0.71$\pm$0.22& 9.70e-21&[2,38,63]\\
		V1108 Her b&47.2$\pm$9.7&0.056854&$\dots$&$147.98^{+2.22}_{-2.16}$&0.88$\pm$0.09& 8.76e-21&[2,29,38]\\
		V592 Her b&38.1$\pm$7.1&0.056097&$\dots$&$710\pm70$&0.6& 5.74e-21&[38,64]\\
		WD 1202-024 b&51.3$\pm$6.3&0.049465&$\dots$&$664.98^{+124.16}_{-90.40}$&0.39$\pm$0.02& 9.01e-21&[2,65]\\
		WD J1820 b&41.9$\pm$31.4&0.065790&$\dots$&$212.62^{+24.00}_{-19.58}$&0.75$\pm$0.18& 4.09e-21&[2,66]\\
		WZ Sge b&60.1$\pm$9.3&0.056688&$\dots$&$45.24\pm0.06$&0.85$\pm$0.04& 1.10e-20&[2,30,38]\\
		ZTF J0003+14 b&17.8$\pm$11.5&0.038540&$\dots$&$247.97^{+38.93}_{-29.63}$&0.79$\pm$0.11& 1.29e-20&[2,67]\\
		ZTF J0220+21 b&14.7$\pm$6.3&0.037150&$\dots$&$340.04^{+71.33}_{-50.25}$&0.83$\pm$0.07& 1.26e-20&[2,67]\\
		ZTF J0407-00 b&19.9$\pm$3.1&0.024583&$\dots$&$747.22^{+287.02}_{-162.32}$&0.79$\pm$0.06& 7.49e-20&[2,67]\\
		ZTF J1637+49 b&24.1$\pm$8.4&0.042710&$\dots$&$204.70^{+8.57}_{-7.91}$&0.9$\pm$0.05& 1.30e-20&[2,67]\\
		ZTF J2252-05 b&27.2$\pm$8.4&0.025970&$\dots$&$511.98^{+97.96}_{-70.85}$&0.76$\pm$0.05& 8.14e-20&[2,67]\\
		Gaia 19cwm b&76.5$\pm$15.7&0.059945&$\dots$&$237.39^{+14.08}_{-12.59}$&0.66$\pm$0.06& 9.46e-21&[2,68]\\
		Gaia14aae b&26.2$\pm$1.4&0.034510&$\dots$&$257.10^{+7.92}_{-7.46}$&0.87$\pm$0.02& 3.03e-20&[2,67,69]\\
		\hline
	\end{longtable}
	\tablefoot{The secondary mass ($m_2$), orbital period ($P_{\rm orb}$), eccentricity ($e$), distance ($D_{L}$), and primary mass ($m_1$) are referenced from the published literature listed in the last column. The $\dot{f}_{\rm gw}$ is computed in this work.}\\
	\tablebib{[1] \cite{2011ApJ...729....8Z}; 
		[2] \cite{2023A&A...674A...1G}; 
		[3] \cite{2009ApJ...699.1113Z}; 
		[4] \cite{2023A&A...675A.199A}; 
		[5] \cite{2021MNRAS.505.5957B}; 
		[6] \cite{2021MNRAS.503.2776Y}; 
		[7] \cite{2008PASP..120..848S}; 
		[8] \cite{2004ApJ...616L.139W}; 
		[9] \cite{2018AJ....156...58B}; 
		[10] \cite{2022A&A...658A..13G}; 
		[11] \cite{1988ApJ...327..732L}; 
		[12] \cite{1996MNRAS.282L..37H}; 
		[13] \cite{2017ApJ...840....9M}; 
		[14] \cite{2004MNRAS.354..355J}; 
		[15] \cite{2018ApJ...858L..13S}; 
		[16] \cite{2017ApJ...836..111K}; 
		[17] \cite{2015ApJ...798..117P}; 
		[18] \cite{2017A&A...598A..34S}; 
		[19] \cite{2015ApJ...814..138B}; 
		[20] \cite{2001ApJ...562..363H}; 
		[21] \cite{2014ApJ...791...67S}; 
		[22] \cite{2023ApJ...956...40F}; 
		[23] \cite{2023ApJ...942....6K}; 
		[24] \cite{2007A&A...465..953I}; 
		[25] \cite{2002ApJ...575L..21M}; 
		[26] \cite{2008MNRAS.383..411P}; 
		[27] \cite{2005A&A...436..647F}; 
		[28] \cite{2022Natur.605...41B}; 
		[29] \cite{2022MNRAS.510.6110P}; 
		[30] \cite{2006MNRAS.373..484K}; 
		[31] \cite{2022MNRAS.509.5086W}; 
		[32] \cite{2018A&A...616A...1G}; 
		[33] \cite{2023MNRAS.523.6114N}; 
		[34] \cite{2024ApJ...963..100K}; 
		[35] \cite{2001ApJ...558L.123G}; 
		[36] \cite{2007ApJ...666.1174R}; 
		[37] \cite{2021MNRAS.508.3877G}; 
		[38] \cite{2024A&A...687A.305M}; 
		[39] \cite{2012MNRAS.425.2548B}; 
		[40] \cite{2020MNRAS.494.3799P}; 
		[41] \cite{2022MNRAS.513.3587Z}; 
		[42] \cite{2024A&A...687A.256C}; 
		[43] \cite{2021ApJ...918...58A}; 
		[44] \cite{2016MNRAS.457.1828K}; 
		[45] \cite{1981ApJ...244..269N}; 
		[46] \cite{2010ApJ...715L.109V}; 
		[47] \cite{2013MNRAS.429..868O}; 
		[48] \cite{2019PASJ...71...48I}; 
		[49] \cite{2004PASP..116.1111S}; 
		[50] \cite{2018MNRAS.481.2523P}; 
		[51] \cite{2011MNRAS.410.1113C}; 
		[52] \cite{2006Sci...314.1578L}; 
		[53] \cite{2008MNRAS.388.1582L}; 
		[54] \cite{2017MNRAS.467.1024M}; 
		[55] \cite{2015A&A...573A..61S}; 
		[56] \cite{2005MNRAS.361..487R}; 
		[57] \cite{2007MNRAS.381..827L}; 
		[58] \cite{2016Natur.533..366H}; 
		[59] \cite{2014MNRAS.437.2894C}; 
		[60] \cite{2021MNRAS.507L..30K}; 
		[61] \cite{2024MNRAS.528..676G}; 
		[62] \cite{2023ApJ...954...63R}; 
		[63] \cite{2013NewA...25....1I}; 
		[64] \cite{1999A&A...342L..45V}; 
		[65] \cite{2017MNRAS.471..976P}; 
		[66] \cite{2025MNRAS.540..633C}; 
		[67] \cite{2022MNRAS.512.5440V}; 
		[68] \cite{2024AstL...50..687K}; 
		[69] \cite{2018MNRAS.476.1663G}.}
}

\begin{landscape}
	\centering
	\setlength{\tabcolsep}{3pt}
	\renewcommand{\arraystretch}{1.38}
	\footnotesize
	\begin{longtable}{lccccccccccccccc}
		\caption{S/N values calculated for various detectors and observation durations.}\label{tab:table_snr}\\
		\hline\hline
		\multicolumn{1}{l}{Source} & \multicolumn{3}{c}{TianQin}  & \multicolumn{3}{c}{LISA}  & \multicolumn{3}{c}{Taiji}  & \multicolumn{3}{c}{DECIGO}  & \multicolumn{3}{c}{BBO} \\
		\cline{2-4}
		\cline{5-7}
		\cline{8-10}
		\cline{11-13}
		\cline{14-16}
		& 2 yr & 4 yr & 8 yr & 2 yr & 4 yr & 8 yr & 2 yr & 4 yr & 8 yr & 2 yr & 4 yr & 8 yr & 2 yr & 4 yr & 8 yr \\
		\hline
		\endfirsthead
		\caption{Continued.}\\
		\hline\hline
		\multicolumn{1}{l}{Source} & \multicolumn{3}{c}{TianQin}  & \multicolumn{3}{c}{LISA}  & \multicolumn{3}{c}{Taiji}  & \multicolumn{3}{c}{DECIGO}  & \multicolumn{3}{c}{BBO} \\
		\cline{2-4}
		\cline{5-7}
		\cline{8-10}
		\cline{11-13}
		\cline{14-16}
		& 2 yr & 4 yr & 8 yr & 2 yr & 4 yr & 8 yr & 2 yr & 4 yr & 8 yr & 2 yr & 4 yr & 8 yr & 2 yr & 4 yr & 8 yr \\
		\hline
		\endhead
		\hline
		\endfoot
		\multicolumn{16}{c}{Source with a PSR primary}\\
		\cline{1-16}
		2S 0918-549 b&$0.57^{+0.22}_{-0.19}$&$0.83^{+0.31}_{-0.27}$&$1.20^{+0.45}_{-0.39}$&$0.74^{+0.28}_{-0.24}$&$2.91^{+1.09}_{-0.94}$&$4.11^{+1.55}_{-1.34}$&$1.13^{+0.43}_{-0.37}$&$2.28^{+0.86}_{-0.74}$&$3.23^{+1.21}_{-1.05}$&$2.56^{+0.96}_{-0.83}$&$4.89^{+1.84}_{-1.59}$&$7.99^{+3.01}_{-2.60}$&$4.83^{+1.82}_{-1.57}$&$8.68^{+3.26}_{-2.82}$&$13.60^{+5.12}_{-4.42}$\\
		4U 0513-40 b&$0.36^{+0.00}_{-0.00}$&$0.53^{+0.01}_{-0.01}$&$0.76^{+0.01}_{-0.01}$&$0.47^{+0.01}_{-0.01}$&$1.90^{+0.02}_{-0.02}$&$2.69^{+0.03}_{-0.03}$&$0.73^{+0.01}_{-0.01}$&$1.49^{+0.02}_{-0.02}$&$2.11^{+0.02}_{-0.02}$&$1.71^{+0.02}_{-0.02}$&$3.19^{+0.04}_{-0.04}$&$5.14^{+0.06}_{-0.06}$&$3.19^{+0.04}_{-0.04}$&$5.59^{+0.06}_{-0.06}$&$8.66^{+0.10}_{-0.10}$\\
		4U 0614+09 b&$0.02^{+0.06}_{-0.01}$&$0.03^{+0.09}_{-0.01}$&$0.05^{+0.13}_{-0.01}$&$0.05^{+0.13}_{-0.01}$&$0.08^{+0.21}_{-0.02}$&$0.11^{+0.30}_{-0.03}$&$0.06^{+0.16}_{-0.02}$&$0.09^{+0.24}_{-0.02}$&$0.12^{+0.33}_{-0.04}$&$0.11^{+0.31}_{-0.03}$&$0.16^{+0.43}_{-0.05}$&$0.23^{+0.61}_{-0.06}$&$0.13^{+0.35}_{-0.04}$&$0.19^{+0.50}_{-0.05}$&$0.27^{+0.71}_{-0.08}$\\
		4U 1543-62 b&$1.02^{+1.39}_{-0.51}$&$1.48^{+2.01}_{-0.74}$&$2.13^{+2.90}_{-1.07}$&$1.30^{+1.77}_{-0.65}$&$4.99^{+6.77}_{-2.49}$&$7.05^{+9.57}_{-3.53}$&$1.98^{+2.69}_{-0.99}$&$3.93^{+5.34}_{-1.97}$&$5.56^{+7.55}_{-2.78}$&$4.21^{+5.72}_{-2.11}$&$8.43^{+11.44}_{-4.21}$&$14.06^{+19.09}_{-7.03}$&$8.07^{+10.96}_{-4.04}$&$15.18^{+20.60}_{-7.59}$&$24.24^{+32.90}_{-12.12}$\\
		4U 1626-67 b&$0.09^{+0.05}_{-0.02}$&$0.13^{+0.07}_{-0.03}$&$0.19^{+0.09}_{-0.05}$&$0.17^{+0.08}_{-0.04}$&$0.29^{+0.14}_{-0.07}$&$0.41^{+0.20}_{-0.10}$&$0.21^{+0.10}_{-0.05}$&$0.31^{+0.15}_{-0.08}$&$0.43^{+0.22}_{-0.11}$&$0.39^{+0.20}_{-0.10}$&$0.56^{+0.28}_{-0.14}$&$0.79^{+0.39}_{-0.20}$&$0.44^{+0.22}_{-0.11}$&$0.62^{+0.31}_{-0.15}$&$0.88^{+0.44}_{-0.22}$\\
		4U 1850-086 b&$0.32^{+0.01}_{-0.01}$&$0.46^{+0.01}_{-0.01}$&$0.66^{+0.02}_{-0.02}$&$0.40^{+0.01}_{-0.01}$&$1.31^{+0.04}_{-0.04}$&$1.86^{+0.06}_{-0.06}$&$0.59^{+0.02}_{-0.02}$&$1.09^{+0.04}_{-0.03}$&$1.54^{+0.05}_{-0.05}$&$0.89^{+0.03}_{-0.03}$&$2.11^{+0.07}_{-0.07}$&$3.96^{+0.13}_{-0.12}$&$1.66^{+0.05}_{-0.05}$&$4.03^{+0.13}_{-0.13}$&$7.20^{+0.23}_{-0.23}$\\
		4U 1905+000 b&$0.01^{+0.00}_{-0.00}$&$0.01^{+0.00}_{-0.00}$&$0.02^{+0.00}_{-0.00}$&$0.02^{+0.00}_{-0.00}$&$0.04^{+0.01}_{-0.00}$&$0.05^{+0.01}_{-0.01}$&$0.03^{+0.00}_{-0.00}$&$0.04^{+0.01}_{-0.01}$&$0.06^{+0.01}_{-0.01}$&$0.06^{+0.01}_{-0.01}$&$0.08^{+0.01}_{-0.01}$&$0.11^{+0.02}_{-0.01}$&$0.07^{+0.01}_{-0.01}$&$0.10^{+0.02}_{-0.01}$&$0.15^{+0.02}_{-0.02}$\\
		IGR J17062-6143 b&$0.04^{+0.01}_{-0.01}$&$0.06^{+0.01}_{-0.01}$&$0.08^{+0.02}_{-0.01}$&$0.07^{+0.01}_{-0.01}$&$0.12^{+0.03}_{-0.02}$&$0.18^{+0.04}_{-0.03}$&$0.09^{+0.02}_{-0.02}$&$0.13^{+0.03}_{-0.02}$&$0.19^{+0.04}_{-0.03}$&$0.17^{+0.03}_{-0.03}$&$0.24^{+0.05}_{-0.04}$&$0.33^{+0.07}_{-0.06}$&$0.18^{+0.04}_{-0.03}$&$0.26^{+0.05}_{-0.04}$&$0.37^{+0.07}_{-0.06}$\\
		M15 X-2 b&$0.15^{+0.00}_{-0.00}$&$0.21^{+0.00}_{-0.00}$&$0.30^{+0.00}_{-0.00}$&$0.19^{+0.00}_{-0.00}$&$0.55^{+0.00}_{-0.00}$&$0.77^{+0.01}_{-0.01}$&$0.27^{+0.00}_{-0.00}$&$0.47^{+0.00}_{-0.00}$&$0.67^{+0.01}_{-0.01}$&$0.44^{+0.00}_{-0.00}$&$0.77^{+0.01}_{-0.01}$&$1.63^{+0.01}_{-0.01}$&$0.48^{+0.00}_{-0.00}$&$1.51^{+0.01}_{-0.01}$&$3.07^{+0.03}_{-0.03}$\\
		MAXI J0911-655 b&$0.02^{+0.00}_{-0.00}$&$0.03^{+0.00}_{-0.00}$&$0.05^{+0.00}_{-0.00}$&$0.04^{+0.00}_{-0.00}$&$0.07^{+0.00}_{-0.00}$&$0.10^{+0.00}_{-0.00}$&$0.05^{+0.00}_{-0.00}$&$0.08^{+0.00}_{-0.00}$&$0.11^{+0.00}_{-0.00}$&$0.10^{+0.00}_{-0.00}$&$0.14^{+0.00}_{-0.00}$&$0.20^{+0.00}_{-0.00}$&$0.11^{+0.00}_{-0.00}$&$0.16^{+0.00}_{-0.00}$&$0.23^{+0.00}_{-0.00}$\\
		NGC 6440 X-2 b&$0.00^{+0.00}_{-0.00}$&$0.01^{+0.00}_{-0.00}$&$0.01^{+0.00}_{-0.00}$&$0.01^{+0.00}_{-0.00}$&$0.01^{+0.00}_{-0.00}$&$0.02^{+0.00}_{-0.00}$&$0.01^{+0.00}_{-0.00}$&$0.01^{+0.00}_{-0.00}$&$0.02^{+0.00}_{-0.00}$&$0.02^{+0.00}_{-0.00}$&$0.03^{+0.00}_{-0.00}$&$0.04^{+0.00}_{-0.00}$&$0.02^{+0.00}_{-0.00}$&$0.03^{+0.00}_{-0.00}$&$0.05^{+0.00}_{-0.00}$\\
		NGC 6652B b&$0.01^{+0.00}_{-0.00}$&$0.02^{+0.00}_{-0.00}$&$0.03^{+0.00}_{-0.00}$&$0.03^{+0.00}_{-0.00}$&$0.05^{+0.00}_{-0.00}$&$0.07^{+0.00}_{-0.00}$&$0.03^{+0.00}_{-0.00}$&$0.05^{+0.00}_{-0.00}$&$0.07^{+0.00}_{-0.00}$&$0.07^{+0.00}_{-0.00}$&$0.09^{+0.00}_{-0.00}$&$0.13^{+0.00}_{-0.00}$&$0.07^{+0.00}_{-0.00}$&$0.10^{+0.00}_{-0.00}$&$0.15^{+0.00}_{-0.00}$\\
		PSR J0636+5129 b&$0.04^{+0.00}_{-0.00}$&$0.06^{+0.01}_{-0.01}$&$0.08^{+0.01}_{-0.01}$&$0.10^{+0.01}_{-0.01}$&$0.15^{+0.02}_{-0.02}$&$0.22^{+0.02}_{-0.03}$&$0.12^{+0.01}_{-0.02}$&$0.18^{+0.02}_{-0.02}$&$0.26^{+0.03}_{-0.03}$&$0.26^{+0.03}_{-0.03}$&$0.36^{+0.04}_{-0.05}$&$0.51^{+0.05}_{-0.06}$&$0.35^{+0.04}_{-0.04}$&$0.49^{+0.05}_{-0.06}$&$0.69^{+0.07}_{-0.09}$\\
		PSR J1311-3430 b&$0.01^{+0.00}_{-0.00}$&$0.01^{+0.00}_{-0.00}$&$0.01^{+0.00}_{-0.00}$&$0.02^{+0.00}_{-0.00}$&$0.02^{+0.00}_{-0.00}$&$0.03^{+0.00}_{-0.00}$&$0.02^{+0.00}_{-0.00}$&$0.03^{+0.00}_{-0.00}$&$0.04^{+0.00}_{-0.00}$&$0.04^{+0.00}_{-0.00}$&$0.05^{+0.01}_{-0.01}$&$0.08^{+0.01}_{-0.01}$&$0.05^{+0.01}_{-0.01}$&$0.07^{+0.01}_{-0.01}$&$0.10^{+0.01}_{-0.01}$\\
		PSR J1653-0158 b&$0.04^{+0.01}_{-0.01}$&$0.06^{+0.01}_{-0.01}$&$0.09^{+0.02}_{-0.01}$&$0.11^{+0.02}_{-0.02}$&$0.16^{+0.03}_{-0.02}$&$0.23^{+0.04}_{-0.03}$&$0.13^{+0.02}_{-0.02}$&$0.19^{+0.04}_{-0.03}$&$0.27^{+0.05}_{-0.04}$&$0.25^{+0.05}_{-0.04}$&$0.35^{+0.07}_{-0.05}$&$0.50^{+0.10}_{-0.08}$&$0.32^{+0.06}_{-0.05}$&$0.45^{+0.09}_{-0.07}$&$0.63^{+0.12}_{-0.10}$\\
		XB 1916-053 b&$0.01^{+0.00}_{-0.00}$&$0.01^{+0.00}_{-0.00}$&$0.02^{+0.01}_{-0.00}$&$0.02^{+0.01}_{-0.00}$&$0.03^{+0.01}_{-0.01}$&$0.05^{+0.01}_{-0.01}$&$0.03^{+0.01}_{-0.00}$&$0.04^{+0.01}_{-0.01}$&$0.05^{+0.01}_{-0.01}$&$0.05^{+0.01}_{-0.01}$&$0.07^{+0.02}_{-0.01}$&$0.10^{+0.02}_{-0.02}$&$0.06^{+0.01}_{-0.01}$&$0.08^{+0.02}_{-0.01}$&$0.11^{+0.03}_{-0.02}$\\
		XTE J1751-305 b&$0.05^{+0.01}_{-0.01}$&$0.07^{+0.02}_{-0.01}$&$0.10^{+0.03}_{-0.02}$&$0.09^{+0.02}_{-0.02}$&$0.15^{+0.04}_{-0.03}$&$0.21^{+0.06}_{-0.04}$&$0.11^{+0.03}_{-0.02}$&$0.16^{+0.04}_{-0.03}$&$0.23^{+0.06}_{-0.05}$&$0.21^{+0.06}_{-0.04}$&$0.29^{+0.08}_{-0.06}$&$0.41^{+0.11}_{-0.09}$&$0.23^{+0.06}_{-0.05}$&$0.33^{+0.09}_{-0.07}$&$0.46^{+0.13}_{-0.10}$\\
		XTE J1807-294 b&$0.03^{+0.06}_{-0.02}$&$0.04^{+0.08}_{-0.03}$&$0.06^{+0.11}_{-0.04}$&$0.05^{+0.10}_{-0.03}$&$0.09^{+0.18}_{-0.06}$&$0.13^{+0.25}_{-0.08}$&$0.07^{+0.13}_{-0.04}$&$0.10^{+0.19}_{-0.06}$&$0.14^{+0.26}_{-0.09}$&$0.13^{+0.24}_{-0.08}$&$0.18^{+0.34}_{-0.11}$&$0.25^{+0.48}_{-0.16}$&$0.14^{+0.27}_{-0.09}$&$0.20^{+0.38}_{-0.13}$&$0.28^{+0.53}_{-0.18}$\\
		ZTF J1406+1222 Ab&$0.08^{+0.02}_{-0.01}$&$0.11^{+0.02}_{-0.02}$&$0.16^{+0.03}_{-0.02}$&$0.20^{+0.04}_{-0.03}$&$0.29^{+0.06}_{-0.04}$&$0.42^{+0.09}_{-0.06}$&$0.23^{+0.05}_{-0.03}$&$0.35^{+0.07}_{-0.05}$&$0.49^{+0.10}_{-0.07}$&$0.46^{+0.10}_{-0.07}$&$0.66^{+0.14}_{-0.10}$&$0.93^{+0.20}_{-0.14}$&$0.59^{+0.13}_{-0.09}$&$0.84^{+0.18}_{-0.13}$&$1.19^{+0.25}_{-0.18}$\\
		\cline{1-16}
		\multicolumn{16}{c}{Source with a WD primary}\\
		\cline{1-16}
		AL Com b&$0.11^{+0.09}_{-0.05}$&$0.15^{+0.13}_{-0.08}$&$0.21^{+0.18}_{-0.11}$&$0.27^{+0.23}_{-0.14}$&$0.40^{+0.34}_{-0.21}$&$0.56^{+0.48}_{-0.29}$&$0.32^{+0.27}_{-0.16}$&$0.47^{+0.40}_{-0.24}$&$0.66^{+0.57}_{-0.35}$&$0.63^{+0.54}_{-0.33}$&$0.90^{+0.77}_{-0.47}$&$1.27^{+1.09}_{-0.66}$&$0.82^{+0.70}_{-0.43}$&$1.16^{+0.99}_{-0.60}$&$1.64^{+1.40}_{-0.85}$\\
		ASASSN-16kr b&$0.26^{+0.02}_{-0.02}$&$0.36^{+0.03}_{-0.02}$&$0.52^{+0.04}_{-0.04}$&$0.67^{+0.05}_{-0.05}$&$0.98^{+0.07}_{-0.07}$&$1.39^{+0.10}_{-0.09}$&$0.79^{+0.06}_{-0.05}$&$1.17^{+0.08}_{-0.08}$&$1.65^{+0.12}_{-0.11}$&$1.61^{+0.11}_{-0.11}$&$2.27^{+0.16}_{-0.15}$&$3.21^{+0.23}_{-0.22}$&$2.12^{+0.15}_{-0.14}$&$3.00^{+0.21}_{-0.20}$&$4.24^{+0.30}_{-0.29}$\\
		ASASSN-17jf b&$0.22^{+0.08}_{-0.06}$&$0.31^{+0.11}_{-0.09}$&$0.44^{+0.16}_{-0.13}$&$0.55^{+0.20}_{-0.17}$&$0.82^{+0.29}_{-0.24}$&$1.16^{+0.41}_{-0.35}$&$0.65^{+0.23}_{-0.19}$&$0.97^{+0.35}_{-0.29}$&$1.37^{+0.49}_{-0.41}$&$1.31^{+0.47}_{-0.39}$&$1.85^{+0.66}_{-0.55}$&$2.62^{+0.94}_{-0.78}$&$1.69^{+0.61}_{-0.51}$&$2.40^{+0.86}_{-0.72}$&$3.39^{+1.21}_{-1.01}$\\
		BW Scl b&$0.68^{+0.09}_{-0.08}$&$0.97^{+0.12}_{-0.12}$&$1.38^{+0.18}_{-0.17}$&$1.73^{+0.22}_{-0.21}$&$2.56^{+0.33}_{-0.31}$&$3.62^{+0.46}_{-0.44}$&$2.04^{+0.26}_{-0.25}$&$3.03^{+0.39}_{-0.37}$&$4.29^{+0.55}_{-0.52}$&$4.06^{+0.52}_{-0.49}$&$5.74^{+0.74}_{-0.70}$&$8.12^{+1.04}_{-0.98}$&$5.19^{+0.67}_{-0.63}$&$7.34^{+0.94}_{-0.89}$&$10.38^{+1.33}_{-1.26}$\\
		CP Eri b&$1.17^{+0.86}_{-0.59}$&$1.65^{+1.22}_{-0.83}$&$2.34^{+1.73}_{-1.18}$&$1.66^{+1.23}_{-0.84}$&$3.64^{+2.69}_{-1.83}$&$5.14^{+3.80}_{-2.59}$&$2.19^{+1.62}_{-1.10}$&$3.49^{+2.58}_{-1.76}$&$4.93^{+3.65}_{-2.48}$&$3.96^{+2.93}_{-2.00}$&$5.60^{+4.14}_{-2.82}$&$7.92^{+5.86}_{-3.99}$&$4.23^{+3.13}_{-2.13}$&$5.98^{+4.42}_{-3.01}$&$14.77^{+10.92}_{-7.44}$\\
		CR Boo b&$5.14^{+2.11}_{-1.79}$&$7.31^{+3.00}_{-2.54}$&$10.41^{+4.27}_{-3.62}$&$6.79^{+2.79}_{-2.36}$&$17.55^{+7.20}_{-6.09}$&$24.82^{+10.19}_{-8.62}$&$9.34^{+3.83}_{-3.24}$&$15.82^{+6.49}_{-5.49}$&$22.37^{+9.18}_{-7.77}$&$15.97^{+6.56}_{-5.55}$&$22.59^{+9.27}_{-7.84}$&$48.70^{+19.99}_{-16.91}$&$16.83^{+6.91}_{-5.84}$&$39.48^{+16.20}_{-13.71}$&$94.34^{+38.72}_{-32.76}$\\
		CRTS J012059.6+325545 b&$0.12^{+0.07}_{-0.05}$&$0.17^{+0.10}_{-0.07}$&$0.25^{+0.14}_{-0.11}$&$0.31^{+0.18}_{-0.13}$&$0.46^{+0.27}_{-0.20}$&$0.65^{+0.38}_{-0.28}$&$0.37^{+0.21}_{-0.16}$&$0.55^{+0.32}_{-0.23}$&$0.77^{+0.45}_{-0.33}$&$0.74^{+0.43}_{-0.32}$&$1.05^{+0.61}_{-0.45}$&$1.48^{+0.86}_{-0.63}$&$0.96^{+0.56}_{-0.41}$&$1.36^{+0.79}_{-0.58}$&$1.92^{+1.11}_{-0.82}$\\
		CRTS J1122-1110 b&$0.04^{+0.05}_{-0.03}$&$0.06^{+0.07}_{-0.04}$&$0.08^{+0.10}_{-0.06}$&$0.10^{+0.12}_{-0.07}$&$0.15^{+0.18}_{-0.11}$&$0.21^{+0.26}_{-0.15}$&$0.12^{+0.14}_{-0.08}$&$0.17^{+0.21}_{-0.12}$&$0.24^{+0.30}_{-0.18}$&$0.22^{+0.28}_{-0.16}$&$0.31^{+0.39}_{-0.23}$&$0.45^{+0.56}_{-0.32}$&$0.27^{+0.34}_{-0.20}$&$0.38^{+0.48}_{-0.28}$&$0.54^{+0.68}_{-0.40}$\\
		DI UMa b&$0.09^{+0.03}_{-0.03}$&$0.13^{+0.04}_{-0.04}$&$0.18^{+0.06}_{-0.05}$&$0.23^{+0.08}_{-0.07}$&$0.34^{+0.11}_{-0.10}$&$0.48^{+0.16}_{-0.14}$&$0.27^{+0.09}_{-0.08}$&$0.40^{+0.13}_{-0.12}$&$0.57^{+0.19}_{-0.16}$&$0.54^{+0.18}_{-0.15}$&$0.76^{+0.25}_{-0.22}$&$1.08^{+0.35}_{-0.31}$&$0.69^{+0.23}_{-0.20}$&$0.98^{+0.32}_{-0.28}$&$1.38^{+0.45}_{-0.40}$\\
		EF Eri b&$0.59^{+0.03}_{-0.03}$&$0.83^{+0.05}_{-0.05}$&$1.18^{+0.07}_{-0.07}$&$1.49^{+0.09}_{-0.09}$&$2.21^{+0.13}_{-0.13}$&$3.13^{+0.18}_{-0.18}$&$1.76^{+0.10}_{-0.10}$&$2.62^{+0.15}_{-0.15}$&$3.70^{+0.22}_{-0.21}$&$3.53^{+0.21}_{-0.20}$&$4.99^{+0.30}_{-0.28}$&$7.06^{+0.42}_{-0.40}$&$4.56^{+0.27}_{-0.26}$&$6.44^{+0.38}_{-0.37}$&$9.11^{+0.54}_{-0.52}$\\
		EG Cnc b&$0.18^{+0.05}_{-0.05}$&$0.26^{+0.07}_{-0.06}$&$0.37^{+0.10}_{-0.09}$&$0.48^{+0.13}_{-0.12}$&$0.70^{+0.20}_{-0.17}$&$0.99^{+0.28}_{-0.24}$&$0.56^{+0.16}_{-0.14}$&$0.83^{+0.23}_{-0.20}$&$1.17^{+0.33}_{-0.29}$&$1.14^{+0.32}_{-0.28}$&$1.61^{+0.45}_{-0.39}$&$2.27^{+0.64}_{-0.56}$&$1.49^{+0.42}_{-0.37}$&$2.11^{+0.59}_{-0.52}$&$2.98^{+0.84}_{-0.73}$\\
		EPIC 212235321 b&$0.17^{+0.04}_{-0.04}$&$0.24^{+0.06}_{-0.05}$&$0.34^{+0.09}_{-0.08}$&$0.41^{+0.11}_{-0.09}$&$0.62^{+0.16}_{-0.14}$&$0.88^{+0.23}_{-0.20}$&$0.49^{+0.13}_{-0.11}$&$0.73^{+0.19}_{-0.16}$&$1.03^{+0.27}_{-0.23}$&$0.95^{+0.25}_{-0.21}$&$1.35^{+0.35}_{-0.30}$&$1.91^{+0.49}_{-0.43}$&$1.18^{+0.31}_{-0.26}$&$1.66^{+0.43}_{-0.37}$&$2.35^{+0.61}_{-0.53}$\\
		EZ Lyn b&$0.29^{+0.11}_{-0.09}$&$0.41^{+0.16}_{-0.13}$&$0.58^{+0.22}_{-0.19}$&$0.75^{+0.29}_{-0.24}$&$1.10^{+0.42}_{-0.36}$&$1.55^{+0.59}_{-0.51}$&$0.88^{+0.33}_{-0.29}$&$1.30^{+0.50}_{-0.42}$&$1.84^{+0.70}_{-0.60}$&$1.78^{+0.68}_{-0.58}$&$2.52^{+0.96}_{-0.82}$&$3.57^{+1.36}_{-1.16}$&$2.34^{+0.89}_{-0.76}$&$3.30^{+1.26}_{-1.07}$&$4.67^{+1.78}_{-1.52}$\\
		GP Com b&$1.26^{+0.88}_{-0.68}$&$1.79^{+1.26}_{-0.97}$&$2.52^{+1.77}_{-1.37}$&$2.46^{+1.73}_{-1.34}$&$4.03^{+2.83}_{-2.19}$&$5.70^{+4.01}_{-3.10}$&$3.06^{+2.15}_{-1.66}$&$4.45^{+3.13}_{-2.42}$&$6.29^{+4.43}_{-3.42}$&$5.76^{+4.06}_{-3.13}$&$8.15^{+5.74}_{-4.43}$&$11.52^{+8.11}_{-6.26}$&$6.57^{+4.63}_{-3.57}$&$9.30^{+6.54}_{-5.05}$&$13.15^{+9.25}_{-7.14}$\\
		GW Lib b&$0.58^{+0.09}_{-0.08}$&$0.82^{+0.12}_{-0.12}$&$1.17^{+0.18}_{-0.16}$&$1.46^{+0.22}_{-0.20}$&$2.17^{+0.32}_{-0.30}$&$3.06^{+0.46}_{-0.43}$&$1.72^{+0.26}_{-0.24}$&$2.56^{+0.38}_{-0.36}$&$3.62^{+0.54}_{-0.51}$&$3.42^{+0.51}_{-0.48}$&$4.83^{+0.72}_{-0.68}$&$6.83^{+1.02}_{-0.96}$&$4.34^{+0.65}_{-0.61}$&$6.14^{+0.92}_{-0.86}$&$8.69^{+1.30}_{-1.22}$\\
		IX Draconis b&$0.05^{+0.03}_{-0.02}$&$0.07^{+0.04}_{-0.03}$&$0.10^{+0.05}_{-0.04}$&$0.13^{+0.07}_{-0.06}$&$0.19^{+0.10}_{-0.08}$&$0.27^{+0.15}_{-0.12}$&$0.15^{+0.08}_{-0.07}$&$0.23^{+0.12}_{-0.10}$&$0.32^{+0.18}_{-0.14}$&$0.32^{+0.17}_{-0.14}$&$0.45^{+0.25}_{-0.20}$&$0.64^{+0.35}_{-0.28}$&$0.43^{+0.23}_{-0.19}$&$0.61^{+0.33}_{-0.26}$&$0.86^{+0.47}_{-0.37}$\\
		NSV 1440 b&$0.73^{+0.02}_{-0.02}$&$1.04^{+0.03}_{-0.03}$&$1.46^{+0.04}_{-0.04}$&$1.22^{+0.03}_{-0.03}$&$2.21^{+0.06}_{-0.06}$&$3.13^{+0.08}_{-0.08}$&$1.55^{+0.04}_{-0.04}$&$2.30^{+0.06}_{-0.06}$&$3.25^{+0.09}_{-0.09}$&$2.90^{+0.08}_{-0.08}$&$4.10^{+0.11}_{-0.11}$&$5.79^{+0.15}_{-0.15}$&$3.18^{+0.08}_{-0.08}$&$4.50^{+0.12}_{-0.12}$&$6.36^{+0.17}_{-0.17}$\\
		OGLE BLG-DN-7 b&$0.16^{+0.04}_{-0.04}$&$0.22^{+0.06}_{-0.06}$&$0.32^{+0.09}_{-0.08}$&$0.41^{+0.11}_{-0.10}$&$0.60^{+0.17}_{-0.15}$&$0.85^{+0.24}_{-0.21}$&$0.48^{+0.13}_{-0.12}$&$0.71^{+0.20}_{-0.18}$&$1.01^{+0.28}_{-0.25}$&$0.97^{+0.27}_{-0.24}$&$1.38^{+0.39}_{-0.34}$&$1.95^{+0.55}_{-0.48}$&$1.28^{+0.36}_{-0.32}$&$1.81^{+0.51}_{-0.45}$&$2.55^{+0.72}_{-0.63}$\\
		OT J1112-3538 b&$0.04^{+0.04}_{-0.02}$&$0.05^{+0.06}_{-0.04}$&$0.08^{+0.08}_{-0.05}$&$0.10^{+0.11}_{-0.06}$&$0.15^{+0.16}_{-0.09}$&$0.21^{+0.23}_{-0.13}$&$0.12^{+0.13}_{-0.07}$&$0.17^{+0.19}_{-0.11}$&$0.24^{+0.27}_{-0.16}$&$0.23^{+0.26}_{-0.15}$&$0.33^{+0.36}_{-0.21}$&$0.47^{+0.51}_{-0.30}$&$0.31^{+0.34}_{-0.20}$&$0.43^{+0.47}_{-0.28}$&$0.61^{+0.67}_{-0.39}$\\
		PQ And b&$0.16^{+0.04}_{-0.04}$&$0.23^{+0.06}_{-0.05}$&$0.33^{+0.09}_{-0.08}$&$0.41^{+0.11}_{-0.10}$&$0.61^{+0.17}_{-0.14}$&$0.86^{+0.23}_{-0.20}$&$0.48^{+0.13}_{-0.11}$&$0.72^{+0.20}_{-0.17}$&$1.02^{+0.28}_{-0.24}$&$0.97^{+0.26}_{-0.23}$&$1.37^{+0.37}_{-0.32}$&$1.94^{+0.53}_{-0.45}$&$1.25^{+0.34}_{-0.29}$&$1.77^{+0.48}_{-0.41}$&$2.50^{+0.68}_{-0.58}$\\
		PR Her b&$0.05^{+0.01}_{-0.01}$&$0.07^{+0.02}_{-0.02}$&$0.10^{+0.03}_{-0.03}$&$0.12^{+0.03}_{-0.03}$&$0.18^{+0.05}_{-0.05}$&$0.26^{+0.07}_{-0.07}$&$0.14^{+0.04}_{-0.04}$&$0.21^{+0.06}_{-0.06}$&$0.30^{+0.08}_{-0.08}$&$0.29^{+0.08}_{-0.08}$&$0.41^{+0.11}_{-0.11}$&$0.57^{+0.15}_{-0.15}$&$0.37^{+0.10}_{-0.10}$&$0.52^{+0.14}_{-0.14}$&$0.73^{+0.19}_{-0.19}$\\
		QZ Lib b&$0.14^{+0.07}_{-0.06}$&$0.19^{+0.10}_{-0.08}$&$0.27^{+0.15}_{-0.11}$&$0.35^{+0.19}_{-0.15}$&$0.52^{+0.28}_{-0.22}$&$0.73^{+0.39}_{-0.31}$&$0.41^{+0.22}_{-0.17}$&$0.61^{+0.33}_{-0.26}$&$0.87^{+0.47}_{-0.36}$&$0.86^{+0.46}_{-0.36}$&$1.21^{+0.65}_{-0.51}$&$1.71^{+0.92}_{-0.72}$&$1.15^{+0.62}_{-0.48}$&$1.62^{+0.87}_{-0.68}$&$2.29^{+1.24}_{-0.96}$\\
		SDSS J0926+3624 b&$1.19^{+0.50}_{-0.43}$&$1.68^{+0.71}_{-0.61}$&$2.39^{+1.00}_{-0.87}$&$1.69^{+0.71}_{-0.61}$&$3.72^{+1.56}_{-1.35}$&$5.26^{+2.21}_{-1.91}$&$2.24^{+0.94}_{-0.81}$&$3.56^{+1.50}_{-1.29}$&$5.03^{+2.12}_{-1.83}$&$4.04^{+1.70}_{-1.47}$&$5.71^{+2.40}_{-2.07}$&$8.08^{+3.40}_{-2.93}$&$4.31^{+1.81}_{-1.56}$&$6.09^{+2.56}_{-2.21}$&$15.18^{+6.39}_{-5.51}$\\
		SDSS J1035+0551 b&$0.32^{+0.03}_{-0.03}$&$0.45^{+0.04}_{-0.04}$&$0.64^{+0.06}_{-0.06}$&$0.81^{+0.08}_{-0.07}$&$1.19^{+0.12}_{-0.11}$&$1.68^{+0.16}_{-0.15}$&$0.95^{+0.09}_{-0.09}$&$1.41^{+0.14}_{-0.13}$&$1.99^{+0.19}_{-0.18}$&$1.91^{+0.18}_{-0.18}$&$2.70^{+0.26}_{-0.25}$&$3.81^{+0.37}_{-0.35}$&$2.47^{+0.24}_{-0.23}$&$3.49^{+0.34}_{-0.32}$&$4.94^{+0.48}_{-0.45}$\\
		SDSS J1057+2759 b&$0.10^{+0.02}_{-0.02}$&$0.15^{+0.03}_{-0.02}$&$0.21^{+0.04}_{-0.03}$&$0.27^{+0.05}_{-0.05}$&$0.40^{+0.07}_{-0.07}$&$0.56^{+0.10}_{-0.09}$&$0.32^{+0.06}_{-0.05}$&$0.47^{+0.08}_{-0.08}$&$0.67^{+0.12}_{-0.11}$&$0.65^{+0.12}_{-0.11}$&$0.92^{+0.17}_{-0.15}$&$1.31^{+0.23}_{-0.22}$&$0.87^{+0.16}_{-0.14}$&$1.23^{+0.22}_{-0.20}$&$1.74^{+0.31}_{-0.29}$\\
		SDSS J1240-0159 b&$0.25^{+0.08}_{-0.08}$&$0.35^{+0.11}_{-0.11}$&$0.49^{+0.15}_{-0.15}$&$0.42^{+0.13}_{-0.13}$&$0.75^{+0.23}_{-0.23}$&$1.05^{+0.32}_{-0.32}$&$0.53^{+0.16}_{-0.16}$&$0.78^{+0.24}_{-0.24}$&$1.10^{+0.34}_{-0.34}$&$0.99^{+0.30}_{-0.30}$&$1.40^{+0.43}_{-0.43}$&$1.98^{+0.61}_{-0.61}$&$1.09^{+0.33}_{-0.33}$&$1.54^{+0.47}_{-0.47}$&$2.18^{+0.67}_{-0.67}$\\
		SDSS J1339+4847 b&$0.23^{+0.01}_{-0.01}$&$0.33^{+0.01}_{-0.01}$&$0.47^{+0.02}_{-0.02}$&$0.60^{+0.03}_{-0.03}$&$0.88^{+0.04}_{-0.04}$&$1.25^{+0.05}_{-0.05}$&$0.70^{+0.03}_{-0.03}$&$1.04^{+0.05}_{-0.04}$&$1.48^{+0.06}_{-0.06}$&$1.42^{+0.06}_{-0.06}$&$2.00^{+0.09}_{-0.09}$&$2.83^{+0.12}_{-0.12}$&$1.83^{+0.08}_{-0.08}$&$2.59^{+0.11}_{-0.11}$&$3.67^{+0.16}_{-0.16}$\\
		SDSS J1507+5230 b&$0.68^{+0.06}_{-0.05}$&$0.98^{+0.09}_{-0.08}$&$1.38^{+0.12}_{-0.11}$&$1.64^{+0.14}_{-0.13}$&$2.48^{+0.22}_{-0.19}$&$3.50^{+0.31}_{-0.27}$&$1.96^{+0.17}_{-0.15}$&$2.90^{+0.26}_{-0.22}$&$4.10^{+0.36}_{-0.32}$&$3.80^{+0.33}_{-0.29}$&$5.38^{+0.47}_{-0.42}$&$7.61^{+0.67}_{-0.59}$&$4.68^{+0.41}_{-0.36}$&$6.62^{+0.58}_{-0.51}$&$9.36^{+0.82}_{-0.73}$\\
		SDSS J1433+1011 b&$0.28^{+0.06}_{-0.05}$&$0.40^{+0.08}_{-0.07}$&$0.57^{+0.12}_{-0.10}$&$0.72^{+0.15}_{-0.13}$&$1.07^{+0.22}_{-0.19}$&$1.51^{+0.30}_{-0.27}$&$0.85^{+0.17}_{-0.15}$&$1.26^{+0.25}_{-0.23}$&$1.78^{+0.36}_{-0.32}$&$1.69^{+0.34}_{-0.31}$&$2.39^{+0.48}_{-0.43}$&$3.38^{+0.68}_{-0.61}$&$2.16^{+0.44}_{-0.39}$&$3.05^{+0.62}_{-0.55}$&$4.31^{+0.87}_{-0.78}$\\
		SDSS J1730+5545 b&$0.04^{+0.01}_{-0.01}$&$0.06^{+0.02}_{-0.02}$&$0.08^{+0.02}_{-0.02}$&$0.06^{+0.02}_{-0.02}$&$0.12^{+0.04}_{-0.04}$&$0.17^{+0.05}_{-0.05}$&$0.08^{+0.03}_{-0.03}$&$0.12^{+0.04}_{-0.04}$&$0.17^{+0.05}_{-0.05}$&$0.15^{+0.05}_{-0.05}$&$0.22^{+0.07}_{-0.07}$&$0.30^{+0.10}_{-0.10}$&$0.17^{+0.05}_{-0.05}$&$0.24^{+0.07}_{-0.07}$&$0.33^{+0.11}_{-0.11}$\\
		SMSS J1606-1000 b&$0.43^{+0.05}_{-0.04}$&$0.61^{+0.06}_{-0.06}$&$0.87^{+0.09}_{-0.09}$&$1.13^{+0.12}_{-0.11}$&$1.66^{+0.18}_{-0.17}$&$2.34^{+0.25}_{-0.24}$&$1.33^{+0.14}_{-0.13}$&$1.97^{+0.21}_{-0.20}$&$2.78^{+0.30}_{-0.28}$&$2.75^{+0.29}_{-0.28}$&$3.89^{+0.41}_{-0.39}$&$5.49^{+0.58}_{-0.55}$&$3.67^{+0.39}_{-0.37}$&$5.19^{+0.55}_{-0.52}$&$7.34^{+0.78}_{-0.74}$\\
		SRGeJ0411 b&$0.09^{+0.01}_{-0.01}$&$0.12^{+0.02}_{-0.02}$&$0.17^{+0.03}_{-0.02}$&$0.23^{+0.03}_{-0.03}$&$0.33^{+0.05}_{-0.05}$&$0.47^{+0.07}_{-0.06}$&$0.27^{+0.04}_{-0.04}$&$0.39^{+0.06}_{-0.05}$&$0.56^{+0.08}_{-0.08}$&$0.56^{+0.08}_{-0.08}$&$0.79^{+0.12}_{-0.11}$&$1.12^{+0.17}_{-0.15}$&$0.76^{+0.11}_{-0.10}$&$1.07^{+0.16}_{-0.15}$&$1.52^{+0.22}_{-0.21}$\\
		SRGeJ0453 b&$0.60^{+0.33}_{-0.26}$&$0.86^{+0.47}_{-0.37}$&$1.21^{+0.66}_{-0.52}$&$1.30^{+0.71}_{-0.56}$&$2.04^{+1.12}_{-0.88}$&$2.89^{+1.58}_{-1.24}$&$1.60^{+0.87}_{-0.69}$&$2.33^{+1.27}_{-1.00}$&$3.30^{+1.80}_{-1.42}$&$3.03^{+1.66}_{-1.30}$&$4.28^{+2.34}_{-1.84}$&$6.05^{+3.31}_{-2.61}$&$3.56^{+1.95}_{-1.53}$&$5.04^{+2.76}_{-2.17}$&$7.13^{+3.90}_{-3.07}$\\
		SSS J0522-3505 b&$0.04^{+0.01}_{-0.01}$&$0.06^{+0.02}_{-0.02}$&$0.08^{+0.02}_{-0.02}$&$0.11^{+0.03}_{-0.03}$&$0.16^{+0.05}_{-0.04}$&$0.22^{+0.07}_{-0.06}$&$0.13^{+0.04}_{-0.04}$&$0.19^{+0.06}_{-0.05}$&$0.27^{+0.08}_{-0.07}$&$0.26^{+0.08}_{-0.07}$&$0.37^{+0.11}_{-0.10}$&$0.52^{+0.15}_{-0.15}$&$0.34^{+0.10}_{-0.10}$&$0.49^{+0.14}_{-0.14}$&$0.69^{+0.20}_{-0.19}$\\
		SW UMa b&$0.35^{+0.11}_{-0.10}$&$0.50^{+0.16}_{-0.14}$&$0.71^{+0.22}_{-0.20}$&$0.90^{+0.28}_{-0.25}$&$1.33^{+0.42}_{-0.36}$&$1.88^{+0.59}_{-0.52}$&$1.06^{+0.33}_{-0.29}$&$1.58^{+0.49}_{-0.43}$&$2.23^{+0.70}_{-0.61}$&$2.13^{+0.66}_{-0.58}$&$3.02^{+0.94}_{-0.83}$&$4.27^{+1.33}_{-1.17}$&$2.76^{+0.86}_{-0.76}$&$3.90^{+1.22}_{-1.07}$&$5.52^{+1.72}_{-1.51}$\\
		V1108 Her b&$0.35^{+0.08}_{-0.08}$&$0.49^{+0.12}_{-0.11}$&$0.70^{+0.17}_{-0.15}$&$0.89^{+0.21}_{-0.19}$&$1.31^{+0.32}_{-0.28}$&$1.86^{+0.45}_{-0.40}$&$1.05^{+0.25}_{-0.23}$&$1.56^{+0.38}_{-0.34}$&$2.20^{+0.53}_{-0.48}$&$2.10^{+0.51}_{-0.46}$&$2.97^{+0.72}_{-0.65}$&$4.21^{+1.02}_{-0.91}$&$2.72^{+0.66}_{-0.59}$&$3.85^{+0.93}_{-0.83}$&$5.44^{+1.32}_{-1.18}$\\
		V592 Her b&$0.05^{+0.02}_{-0.01}$&$0.07^{+0.02}_{-0.02}$&$0.09^{+0.03}_{-0.02}$&$0.12^{+0.04}_{-0.03}$&$0.18^{+0.06}_{-0.04}$&$0.25^{+0.08}_{-0.06}$&$0.14^{+0.04}_{-0.04}$&$0.21^{+0.07}_{-0.05}$&$0.30^{+0.09}_{-0.07}$&$0.28^{+0.09}_{-0.07}$&$0.40^{+0.13}_{-0.10}$&$0.56^{+0.18}_{-0.14}$&$0.36^{+0.12}_{-0.09}$&$0.51^{+0.16}_{-0.13}$&$0.73^{+0.23}_{-0.18}$\\
		WD 1202-024 b&$0.07^{+0.02}_{-0.02}$&$0.10^{+0.03}_{-0.03}$&$0.14^{+0.04}_{-0.04}$&$0.17^{+0.05}_{-0.04}$&$0.25^{+0.08}_{-0.07}$&$0.36^{+0.11}_{-0.09}$&$0.20^{+0.06}_{-0.05}$&$0.30^{+0.09}_{-0.08}$&$0.42^{+0.13}_{-0.11}$&$0.40^{+0.12}_{-0.10}$&$0.56^{+0.17}_{-0.14}$&$0.79^{+0.24}_{-0.20}$&$0.49^{+0.15}_{-0.13}$&$0.70^{+0.21}_{-0.18}$&$0.99^{+0.30}_{-0.26}$\\
		WD J1820 b&$0.13^{+0.14}_{-0.09}$&$0.18^{+0.19}_{-0.12}$&$0.26^{+0.28}_{-0.18}$&$0.34^{+0.36}_{-0.23}$&$0.50^{+0.52}_{-0.33}$&$0.70^{+0.74}_{-0.47}$&$0.40^{+0.42}_{-0.27}$&$0.59^{+0.62}_{-0.40}$&$0.84^{+0.88}_{-0.56}$&$0.83^{+0.88}_{-0.56}$&$1.18^{+1.24}_{-0.79}$&$1.67^{+1.76}_{-1.12}$&$1.12^{+1.18}_{-0.75}$&$1.59^{+1.67}_{-1.07}$&$2.25^{+2.37}_{-1.51}$\\
		WZ Sge b&$1.42^{+0.23}_{-0.22}$&$2.01^{+0.32}_{-0.30}$&$2.86^{+0.46}_{-0.43}$&$3.62^{+0.59}_{-0.55}$&$5.35^{+0.86}_{-0.81}$&$7.57^{+1.22}_{-1.15}$&$4.26^{+0.69}_{-0.65}$&$6.34^{+1.02}_{-0.96}$&$8.97^{+1.45}_{-1.36}$&$8.57^{+1.38}_{-1.30}$&$12.12^{+1.96}_{-1.84}$&$17.14^{+2.77}_{-2.60}$&$11.08^{+1.79}_{-1.68}$&$15.66^{+2.53}_{-2.37}$&$22.15^{+3.58}_{-3.36}$\\
		ZTF J0003+14 b&$0.21^{+0.20}_{-0.13}$&$0.30^{+0.29}_{-0.18}$&$0.42^{+0.40}_{-0.26}$&$0.45^{+0.44}_{-0.28}$&$0.71^{+0.68}_{-0.44}$&$1.00^{+0.96}_{-0.62}$&$0.55^{+0.53}_{-0.34}$&$0.81^{+0.78}_{-0.50}$&$1.14^{+1.10}_{-0.71}$&$1.05^{+1.01}_{-0.65}$&$1.48^{+1.43}_{-0.92}$&$2.10^{+2.02}_{-1.30}$&$1.24^{+1.19}_{-0.77}$&$1.75^{+1.69}_{-1.08}$&$2.47^{+2.39}_{-1.53}$\\
		ZTF J0220+21 b&$0.14^{+0.10}_{-0.07}$&$0.20^{+0.15}_{-0.10}$&$0.29^{+0.21}_{-0.14}$&$0.30^{+0.22}_{-0.15}$&$0.48^{+0.34}_{-0.24}$&$0.68^{+0.49}_{-0.34}$&$0.37^{+0.27}_{-0.18}$&$0.54^{+0.39}_{-0.27}$&$0.77^{+0.55}_{-0.38}$&$0.70^{+0.51}_{-0.35}$&$1.00^{+0.72}_{-0.50}$&$1.41^{+1.01}_{-0.70}$&$0.82^{+0.59}_{-0.41}$&$1.17^{+0.84}_{-0.58}$&$1.65^{+1.19}_{-0.82}$\\
		ZTF J0407-00 b&$0.25^{+0.13}_{-0.10}$&$0.36^{+0.18}_{-0.14}$&$0.51^{+0.25}_{-0.20}$&$0.41^{+0.21}_{-0.16}$&$0.76^{+0.38}_{-0.30}$&$1.08^{+0.53}_{-0.42}$&$0.53^{+0.26}_{-0.21}$&$0.79^{+0.39}_{-0.31}$&$1.11^{+0.55}_{-0.44}$&$0.99^{+0.49}_{-0.39}$&$1.40^{+0.69}_{-0.55}$&$1.97^{+0.98}_{-0.77}$&$1.08^{+0.53}_{-0.42}$&$1.53^{+0.76}_{-0.60}$&$2.16^{+1.07}_{-0.85}$\\
		ZTF J1637+49 b&$0.28^{+0.12}_{-0.10}$&$0.40^{+0.17}_{-0.14}$&$0.57^{+0.24}_{-0.20}$&$0.65^{+0.28}_{-0.23}$&$0.99^{+0.42}_{-0.35}$&$1.40^{+0.60}_{-0.49}$&$0.78^{+0.33}_{-0.27}$&$1.15^{+0.49}_{-0.40}$&$1.63^{+0.69}_{-0.57}$&$1.50^{+0.64}_{-0.52}$&$2.12^{+0.90}_{-0.74}$&$3.00^{+1.28}_{-1.05}$&$1.81^{+0.77}_{-0.63}$&$2.56^{+1.09}_{-0.89}$&$3.61^{+1.54}_{-1.26}$\\
		ZTF J2252-05 b&$0.43^{+0.23}_{-0.17}$&$0.60^{+0.32}_{-0.24}$&$0.85^{+0.46}_{-0.34}$&$0.72^{+0.39}_{-0.29}$&$1.29^{+0.70}_{-0.52}$&$1.83^{+0.98}_{-0.73}$&$0.91^{+0.49}_{-0.37}$&$1.35^{+0.73}_{-0.54}$&$1.91^{+1.03}_{-0.77}$&$1.71^{+0.92}_{-0.69}$&$2.42^{+1.31}_{-0.98}$&$3.43^{+1.85}_{-1.38}$&$1.89^{+1.02}_{-0.76}$&$2.67^{+1.44}_{-1.07}$&$3.78^{+2.04}_{-1.52}$\\
		Gaia 19cwm b&$0.25^{+0.07}_{-0.06}$&$0.35^{+0.10}_{-0.09}$&$0.50^{+0.14}_{-0.12}$&$0.64^{+0.18}_{-0.16}$&$0.94^{+0.27}_{-0.23}$&$1.33^{+0.38}_{-0.32}$&$0.75^{+0.21}_{-0.18}$&$1.11^{+0.32}_{-0.27}$&$1.57^{+0.45}_{-0.38}$&$1.52^{+0.43}_{-0.37}$&$2.16^{+0.62}_{-0.53}$&$3.05^{+0.87}_{-0.75}$&$2.00^{+0.57}_{-0.49}$&$2.83^{+0.81}_{-0.69}$&$4.00^{+1.14}_{-0.98}$\\
		Gaia14aae b&$0.42^{+0.04}_{-0.03}$&$0.60^{+0.05}_{-0.05}$&$0.84^{+0.07}_{-0.07}$&$0.86^{+0.07}_{-0.07}$&$1.38^{+0.12}_{-0.11}$&$1.95^{+0.17}_{-0.16}$&$1.06^{+0.09}_{-0.09}$&$1.54^{+0.13}_{-0.13}$&$2.18^{+0.19}_{-0.18}$&$2.00^{+0.17}_{-0.16}$&$2.83^{+0.24}_{-0.23}$&$4.00^{+0.34}_{-0.33}$&$2.31^{+0.20}_{-0.19}$&$3.26^{+0.28}_{-0.27}$&$4.61^{+0.40}_{-0.38}$\\
	\end{longtable}
\end{landscape}

\FloatBarrier
\clearpage

\end{appendix}

\end{document}